\documentclass[%
 reprint,
 amsmath,amssymb,
 aps,
 prl,
]{revtex4-2}

\usepackage{dcolumn}

\usepackage{amsmath,amsthm,amssymb}
\usepackage[dvipsnames]{xcolor} 
\usepackage{lmodern} 
\usepackage{graphicx}
\usepackage{float} 
\usepackage{bm} 
\usepackage{braket}
\usepackage{slashed} 
\usepackage{tikz}
\usepackage[compat=1.1.0]{tikz-feynman}
\usepackage[utf8]{inputenc}
\usepackage{xr}

\renewcommand{\t}[1]{\text{#1}}

\newcommand*{\iu}{{\mathrm{i}\mkern1mu}}
\def\bfr{{\bf r}}

\newcommand{\td}{\t{d}}

\newcommand{\Eh}{\t{E}_{\t{h}}}

\usepackage{filecontents}

\DeclareUnicodeCharacter{2212}{-}
\begin{document}

\preprint{APS/123-QED}

\title{Explaining the magnitude of Chirality-Induced Spin Selectivity via electron-electron exchange}
\author{Bence Csakany}
 \email{bc528@cam.ac.uk}
\author{Alex J. W. Thom}
\email{ajwt3@cam.ac.uk}
 \affiliation{Yusuf Hamied Department of Chemistry, University of Cambridge, Lensfield Road, Cambridge, UK.  CB1 2EW}

\begin{abstract}
Chiral molecular structure can couple to electron spin, leading to unexpected spin polarization effects. This Chirality-Induced Spin Selectivity (CISS) was first reported for DNA molecules adsorbed on gold but its microscopic origin remains unclear. We demonstrated though simulation that the exchange arising from electron-electron Coulomb interactions within the self-consistent mean field (Hartree--Fock) approximation can yield significant ($\sim 2\%$) spin polarisation for $3$-methylcyclohexanone adsorbed on Cu(111) amplifying a much smaller ($\sim 0.0014\%$) initial bias, consistent with experiment. Symmetry considerations ensure the result is physically meaningful, while its ab-initio nature ensures all parameters are physically realistic. This amplification is connected to existing studies on spin-symmetry breaking in Hartree--Fock, providing a new pathway for understanding the magnitude of CISS as an emergent phenomenon of interacting electrons. 
\end{abstract}

\maketitle

\section{Introduction}

Chirality-Induced Spin Selectivity (CISS) is a phenomenon in which structural chirality leads to strong spin-dependent electronic behaviour. This is in spite of the weakness of the relativistic spin-orbit coupling which is believed to induce it. CISS has been observed using a diverse range of molecules such as DNA, helicenes, oligopeptides, and chiral self-assembled monolayers\cite{gohler_spin_2011,safari_enantioselective_2024,giaconi_efficient_2023}. The effect is typically measured by electron transmission experiments; however, CISS also has an effect on rates of reaction\cite{lu_enantiospecificity_2021}. Beyond being of intrinsic interest, CISS has also been suggested as an explanation of the homochirality of life\cite{ozturk_origin_2023} as well as having applications in spintronics\cite{michaeli_new_2017}. \\

Here we focus on CISS as observed in UV photo-emission spectroscopy experiments, where the spin polarisation (which can range from 0\% (unpolarized) to 100\% (fully spin polarized)) of the resulting photoelectron current is observed in the 1--60\% range, depending on the system. 
Despite its experimental reporting over a decade ago\cite{gohler_spin_2011}, the mechanism underlying the CISS effect is still uncertain\cite{bloom_chiral_2024,evers_theory_2022}. Various microscopic mechanisms have been proposed\cite{fransson_chiral_2024,dalum_theory_2019,shitade_geometric_2020,zheng_chirality-driven_2025,di_ventra_chirality-induced_2025}, though none have yet produced a predictive and generally accepted explanation for the magnitude of the effect. Though many explanations reproduce a large ($\sim 10\%$) photoelectron CISS effect, this is typically caused by invoking unphysical parameters\cite{fransson_chiral_2025,upadhyay_weak_2026} or accidental degeneracy\cite{dalum_theory_2019}. \\

A leading explanation for the CISS effect originates in electron--electron exchange\cite{ozturk_origin_2023,fransson_chirality-induced_2019}, which modifies the Coulombic repulsion between electrons of the same spin. However, these Coulomb interactions do not, by themselves, connect \textit{chirality} to spin. In this work, we show that Coulombic electron-electron exchange interactions create an amplification of an existing spin bias and this correctly predicts the magnitude of the observed spin polarisation seen in the the CISS effect. It requires no unphysical parameters, accidental degeneracy or fine-tuning.  The exchange interaction provides the amplification necessary for relativistic and chiral phonon effects to create the experimentally observed magnitudes of the CISS effect in a wide range of structures. In contrast to other ab-initio studies\cite{naskar_chiral-induced_2023} incorporating exchange interactions as well as other effects, we show that the exchange interaction is the primary cause of the magnification.

\section{Background}

\begin{figure}[h!]
\centering
\includegraphics[clip, trim=3.5cm 4cm 1.5cm 4cm, width=\linewidth]{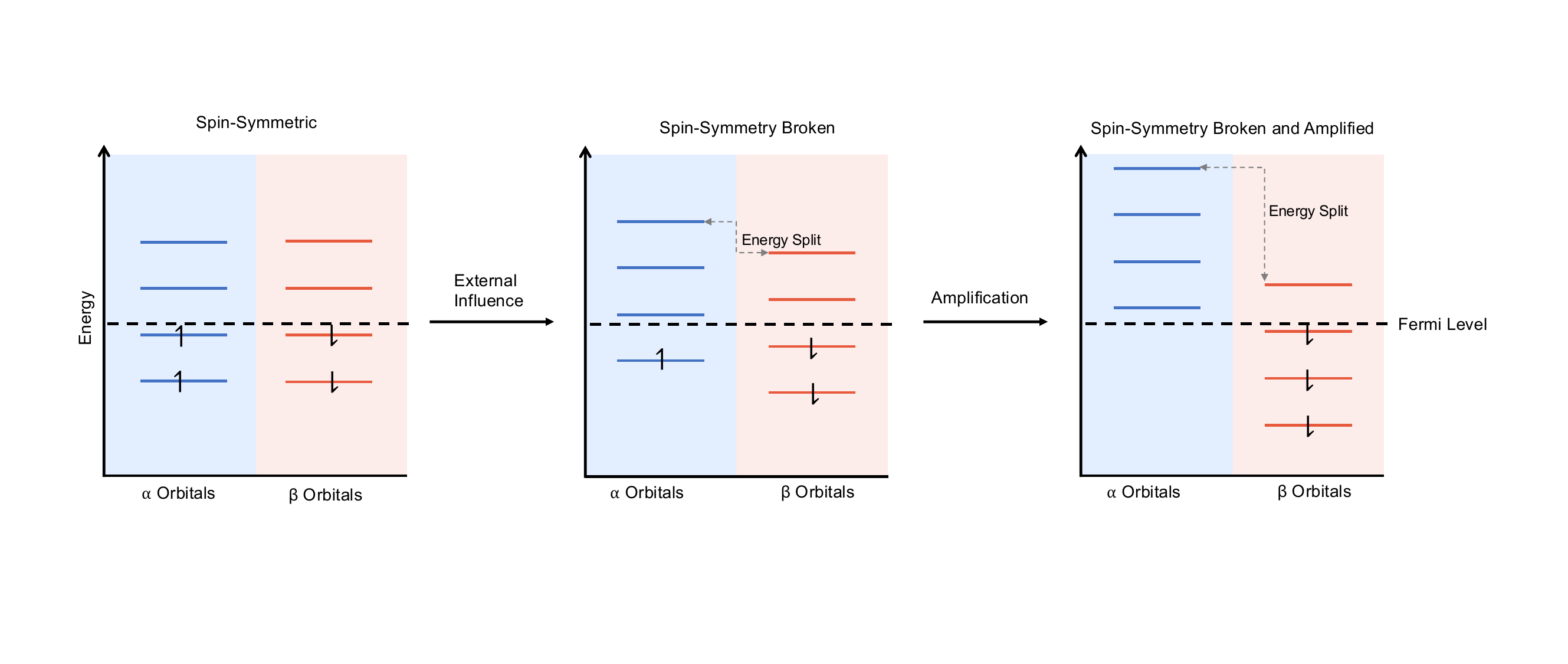}
\caption{\textbf{Schematic diagram showing the change in energy of each electron in a molecule due to an external influence.} Each electron is in an approximate single particle state, a molecular orbital. The energy of these orbitals is perturbed slightly by an external effect such as chiral phonons or spin-orbit coupling, along with breaking of time-reversal symmetry. Once the symmetry is broken, electron--electron exchange effects amplify the symmetry breaking leading to a much larger effect.}
\label{fig:MoDiagram}
\end{figure}

Spin-Orbit Coupling (SOC) is responsible for coupling the chirality of molecular geometry to spin. SOC results from the coupling of the momentum and the spin of electrons in relativistic (Dirac) theory, along with the conversion of an electric field into a magnetic field via Lorentz transformations.  Even with SOC, point group and time reversal symmetries can prevent a spin polarisation being observed (see SI) and so CISS is always observed in non-equilibrium scenarios, with molecules that have no significant point group symmetry.  A detailed understanding of SOC is not necessary here; it is sufficient to understand that for electrons with a net (possibly angular) momentum in an electric field, typically there is a small energetic difference between up and down spin. This energy difference is shown as an `external influence' in figure \ref{fig:MoDiagram}. \\

Dalum\cite{dalum_theory_2019} showed that in the case of orbitals close in energy, a significant ($\sim$10--20\%) spin polarisation can arise using SOC. However when the orbitals were further in energy only a much smaller polarisation was obtained. This is not an issue here however, as we show only a small spin polarisation is needed for the exchange effects (`amplification' in figure \ref{fig:MoDiagram}) to cause amplification.

Another source of small energetic differences between up and down spin electrons is from chiral phonons. Chiral phonons are molecular/crystal nuclear vibrations that have angular momentum. It has been demonstrated in both experiment and theory that chiral phonons can couple to electronic degrees of freedom in a spin-selective way \cite{fransson_vibrational_2020,luo_large_2023}. When out of thermal equilibrium, these chiral phonons act as a spin-pump creating a buildup of spin on an interface\cite{kim_chiral-phonon-activated_2023}. This causes the small difference in the energy of up/down spin as shown in figure \ref{fig:MoDiagram}, which is then amplified by exchange effects. In thermal equilibrium, however, no such effect is observed due to time-reversal symmetry\cite{dalum_theory_2019}. 
Point group symmetries can also prevent a spin polarisation as shown in SI-\ref{SI-sec:SymmetryConsiderations}.
Extensive theoretical studies have attempted to use chiral phonons to predict a macroscopic spin difference\cite{fransson_charge_2021,fransson_charge_2022,fransson_chiral_2024}. However, to achieve this in current models purely via phonons, it appears that unphysically small phonon frequencies must be assumed.\\

Ab-initio\cite{naskar_chiral-induced_2023,zheng_chirality-driven_2025} and model\cite{zollner_insight_2020,dalum_theory_2019,shitade_geometric_2020,medina_continuum_2015} Hamiltonians have both been used to investigate the effect of electron-electron and electron-nucleus spin-orbit coupling. Both within the fully relativistic and perturbative relativistic pictures, these calculations have shown the existence of solutions with a spin texture. Because these were complicated relativistic calculations, the effects of exchange, if included, and spin-orbit coupling were not explicitly disentangled and how the spin texture came about was not the focus. It is this disentanglement of effects that is performed here.\\

\section{Model Experiments}
To computationally investigate the CISS effect in photoelectron spectroscopy we have created two simulated experiements based upon a chiral molecule, $3$-methylcyclohexanone ($3$-MCHO), adsorbed to a Cu(111) surface as shown in figure \ref{fig:NonEqDensityPIAndPZ}. This system is a simplified model of the experiment conducted in reference \citenum{badala_viswanatha_vectorial_2022} where the spin- and energy-resolved photoelectron spectrum for $3$-MCHO adsorbed on Cu(643) was obtained.

The simulations are performed in a Green's function framework consisting of semi-infinite modelled leads and an explicitly represented \textbf{system}.

The first `equilibrium' experiment is partitioned into two sections: i) a \textbf{bulk} copper lead; and ii) the \textbf{system} consisting of a single layer of Cu(111) with 3-MCHO adsorbed onto it.
  The \textbf{bulk} copper lead is modelled by a electron sea via its Green's function.  This couples to the \textbf{system} whose is explicitly represented in a localised Gaussian basis and described by a conventional quantum chemical Hamiltonian.

The second `photo-electron' experiment contains the same components as the `equilibrium', but with two additional leads: iii) the \textbf{photoelectron} lead is a semi-infinite lead, physically superimposed on the \textbf{bulk} lead, but does not interact with it; and iv) a \textbf{vacuum} lead lying beyond the \textbf{system} to act as a sink of photoelectrons.  The \textbf{photoelectron} lead represents a sea of excited photoelectrons with a lower occupied density of states and higher Fermi energy than the bulk, via a different Green's function, which also coupled to the \textbf{system}.  The \textbf{vacuum} lead is represented by a further Green's function with a lower density of states and a lower Fermi energy to act as a sink of outgoing photoelectrons.

\begin{figure}[h!]
\centering
\includegraphics[width=\linewidth]{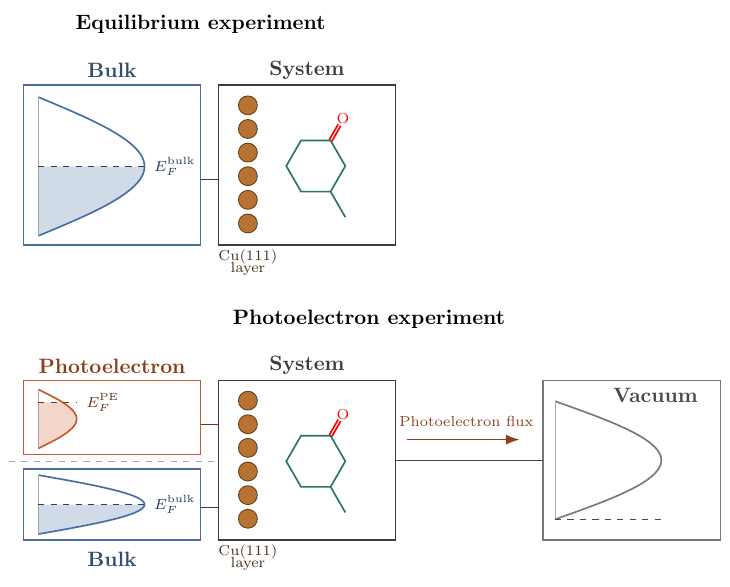}
\caption{Schematic of the Green's-function models used to study photoelectron transport through 3-methylcyclohexanone (3-MCHO) adsorbed on Cu(111). Top: `equilibrium' setup comprising a semi-infinite bulk Cu lead coupled to the explicit Cu(111)$+$3-MCHO system. Bottom: `photoelectron' setup, where additional photoelectron and vacuum leads are introduced to model excited-electron injection and photoelectron emission, respectively. Shaded regions indicate occupied states below the corresponding Fermi levels.}
\label{fig:ModelExpFigPhotoelectron}
\end{figure}

These are depicted in Figure \ref{fig:ModelExpFigPhotoelectron}.

\section{Modelling Non-Equilibrium Transport}
The time evolution of a non-relativistic quantum system is given by the time-dependent Schr\"{o}dinger equation. 

\begin{equation}
\hat{H} \psi = \iu \partial_t \psi
\end{equation}

To model the \textbf{system} we will work in second quantisation in a discrete basis and use the Born--Oppenheimer approximation. Creation/annihilation operators on site $p$ are denoted $\hat a_p^\dagger/\hat a_p$ respectively. $\hat{\mathcal{H}}$ denotes the electronic Hamiltonian and is:

\begin{equation}
\hat{\mathcal{H}} = h_{pq}\hat a^\dagger_p \hat a_q + h_{pqrs} \hat a^\dagger_p \hat a^\dagger_q \hat a_r \hat a_s
\end{equation}
$h_{pq}$ are the one-electron interactions such as momentum and nuclear attraction. $h_{pqrs}$ are the two electron interactions --- the Coulomb interaction of electrons with each other. \\
 
To describe time-dependent non-equilibrium phenomena, it is helpful to use the Green's function, a formally exact encoding of the one particle properties of the system described by this Hamiltonian. Within many-body perturbation theory, the Hartree--Fock (HF) approximation of the Green's function leads to the effective one-particle Hamiltonian:

\begin{equation}
\hat{H} = h_{pq}\hat a^\dagger_p \hat a_q + \frac{1}{2\pi}(h_{pqrs}-h_{qprs}) G^n_{rp} \hat a^\dagger_q \hat a_s\label{eq:HFApproximation}
\end{equation}
where $G^n$ is the (possibly non-)equilibrium density matrix (equation \ref{eq:GnDefinition}) and the two-electron Coulomb integrals are given by:
\begin{equation}
h_{pqrs} = \braket{q(\bfr_1)p(\bfr_2)|\frac1{\lvert \bfr_1-\bfr_2\rvert} |r(\bfr_2)s(\bfr_1)}.
\end{equation}
Within electronic structure theory, the effective one particle Hamiltonian $\hat H$ is called the Fock Hamiltonian.

The framework of Green's functions has the additional advantage that it can also describe an infinite sea of electrons, allowing easy modelling of the leads in our experiment.
The details of non-equilibrium Green's functions(NEGF) are found in standard texts\cite{datta_electronic_1995,rammer_quantum_2007}, and further details are given in the supplementary information (see SI-\ref{SI-SI:NonOrthGreensFuncSec}).  The system Hamiltonian, $\hat H$ defines the Dyson equations below, which are solved to produce Green's functions ($G$) and self-energies ($\Sigma$) describing the system.  The properties of the leads are specified by their self energies ($\Sigma$) which are given later.

For completeness, the relevant Dyson equations are:
\begin{align}
(EI  + \iu \eta - &\hat H - \Sigma^R ) G^R(E) = I \t{, } \eta > 0\\
(EI  - \iu \eta - &\hat H -\Sigma^A) G^A(E) = I\t{, } \eta > 0\\
G^K &= G^R(E) \Sigma^K G^A(E)\\
G^n &= \int_{-\infty}^{\infty} -\iu G^<(E) \td E \label{eq:GnDefinition}\\
G^<(E) &= G^K(E) - G^A(E) - G^R(E)
\end{align}
These are solved to yield the spectral Green's functions ($G^{R/A}$), which describe the possible one particle states of the system, and the Keldysh $G^K$ which describe the occupied one particle states. The self energies $\Sigma^{R/A/K}$ are used in the electron transport calculation to describe reservoirs that can source/sink electrons. The molecular orbitals (MOs) of electronic structure theory correspond to the residue of the divergences of $G^{R/A}$ (i.e. the long lived single particle states). \\

\section{Model Parameters}
As discussed above, for a transport calculation reservoirs (leads) from which electrons can enter/leave the system of interest must be defined. Each reservoir is modelled as an infinite band with a uniform density of states -- often termed the wide-band limit, (see SI-\ref{SI-SI:SelfEnergyTerms}) as it approximates the situation when the range of energy levels of interest are much closer together than the band they are connected to.
To this effect each lead self energy is given by:

\begin{align}
\Sigma^R_{\t{lead } i} &= -\iu \frac{\gamma_i}{2} C_i\\
\Sigma^A_{\t{lead } i} &= \iu \frac{\gamma_i}{2} C_i\\
\Sigma^K_{\t{lead } i} &= -\iu \left(1-2f_i(E)\right)\frac{\gamma_i}{2}C_i
\end{align}
$C_i$ is the coupling matrix that dictates the coupling to each explicit basis function (denoted $i$) in the system. For the self energy representing the \textbf{bulk} copper, $C$ is chosen such that the it couples only to the copper atoms. The \textbf{vacuum} consists of outgoing photoelectrons plane wave states that couple only to the basis functions on the adsorbed molecule. This is discussed in more detail in SI-\ref{SI-SI:SelfEnergyTerms} and shown schematically in figure \ref{fig:ModelExpFigPhotoelectron}.
 $f_i$ is the Fermi function with chemical potential $\mu_i$. $\gamma_i$ is a parameter for each lead that encompasses both the density of states and the strength of the coupling of the system to the lead. An approximation for $\gamma$ is 
\begin{equation}
\gamma = 2\pi t^2 \left(\t{D.O.S}\right)
\end{equation}
Where $t$ is the interaction energy between the two systems, $\t{D.O.S}$ is the density of states of the copper surface. 
The $\gamma$ parameters are chosen to be physically meaningful. The value for the bulk copper lead is:
\begin{equation}
\gamma_{\t{bulk}} = \gamma_{\t{vacuum}} = 10^{-2} \t{E}_{\t{h}}
\end{equation}
To represent the reduced occupation number of the photoelectrons the $\gamma$ term there is reduced:
\begin{equation}
\gamma_{\t{photoelectron}} = 10^{-4}\t{E}_{\t{h}}.
\end{equation}
The results are qualitatively the same (section SI-\ref{SI-SI:sec:DifferentParameters}) independent of the exact value chosen.\\

The chemical potential of the bulk copper was set at $\mu_{\t{bulk}} = -0.15 \t{E}_{\t{h}} = -4.08 \t{eV}$ in order to create a neutral system. The experimental\cite{anderson_work_1949} work function for copper is $-4.46\t{eV}$. This discrepancy of $0.38\t{eV}$ is explained by the approximations made. \\

\section{Results}
In the wavefunction found in ground state electronic structure theory, if the MOs of $\alpha/\beta$ electrons are different, then $\alpha/\beta$ electrons will have different transport properties. This is exactly the CISS effect: a difference in transport properties of $\alpha/\beta$ electrons.
In this non-equilibrium system, the Green's function takes the place of the wavefunction describing the \textbf{system}.   Here we investigate the effect that a small spin perturbation in the bulk copper has on the Hartree--Fock Green's function of the system and the resulting photoelectron spin current.

We are interested in measuring net spin polarisation.  For quantity $Q$, the spin polarisation is given by
\begin{equation}
\t{Spin Polarisation} = \frac{Q_{\uparrow} - Q_{\downarrow}}{Q_{\uparrow} + Q_{\downarrow}} \label{eq:SpinPolEq}
\end{equation}
where $Q_{\uparrow}$ is the quantity of interest in the `up' spin direction and $Q_{\downarrow}$ in the `down' spin direction.

In order induce a spin polarisation in the copper surface as seen in the experiment \cite{badala_viswanatha_vectorial_2022}, $\gamma_{\t{bulk}}$ and $\gamma_{\t{photoelectron}}$ were made spin dependent by increasing the $\gamma$ for $\alpha$ electrons by a factor $1.04$ and decreasing $\gamma$ for $\beta$ electrons by a factor $0.96$\footnote{This choice is arbitrary, reversing the $\alpha$ and $\beta$ electrons would yield related results as seen in CISS experiments where the $L$ enantiomer is exchanged for $R$, SI-\ref{SI-sec:SymmetryConsiderations} }. This method has been used elsewhere to represent a magnetised lead\cite{fransson_charge_2021,dalum_theory_2019}. We do not agree with this interpretation however, as a magnetised lead would have a significant exchange effect on the system which this simple method neglects. This perturbation has been introduced to model the spin symmetry breaking as would be caused by spin-orbit coupling or chiral phonons from the molecule--surface system. \\

No such perturbation was applied to the outgoing photoelectron lead, $\gamma_{\t{vacuum}}$. It is stressed that modifying $\gamma_{\t{photoelectron}}$ in this way does not result in a $4\%$ polarisation of the electron density. Indeed the net spin polarisation of the density (equation \ref{eq:SpinPolEq})  in the copper in `equilibrium' is only $0.0014\%$. This is because the change in $\gamma$ only effects those states that are within $10 k_{\t{b}} T \approx 10\t{mE}_{\t{h}}$ of the Fermi level. All lower states are fully occupied and a change in broadening has no effect.\\

To fit with experiment the chemical potential of the incoming photoelectrons was set at $-0.15\Eh + 5.9\t{eV}$. This corresponds to a $210\t{nm}$ UV light source. \\
\begin{figure}[h!]
\centering
\includegraphics[clip, trim=1.5cm 1.5cm 1.5cm 1.5cm, width=0.3\linewidth]{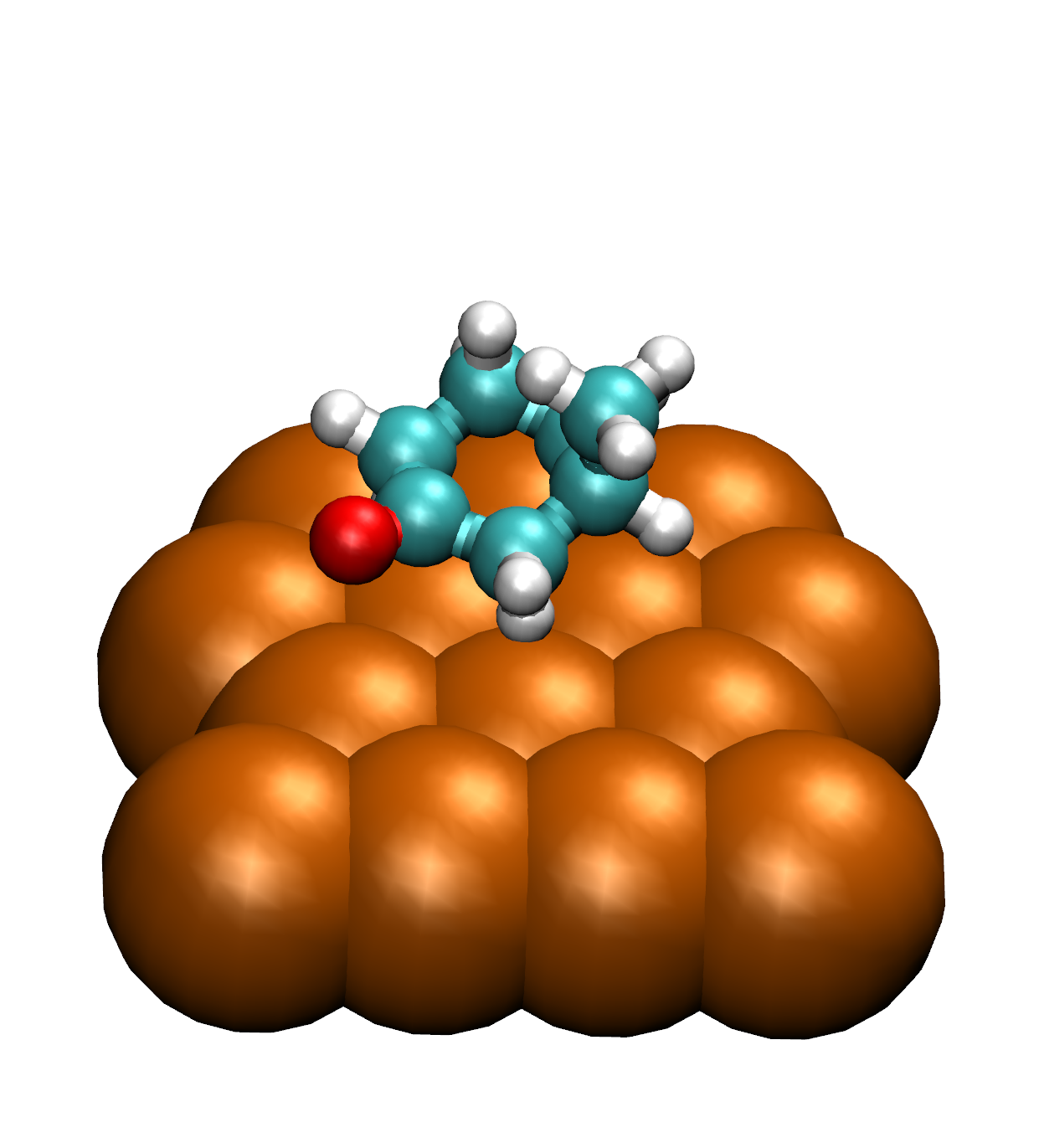}
\includegraphics[clip, trim=3.5cm 3.5cm 3.5cm 3.5cm,width=0.3\linewidth]{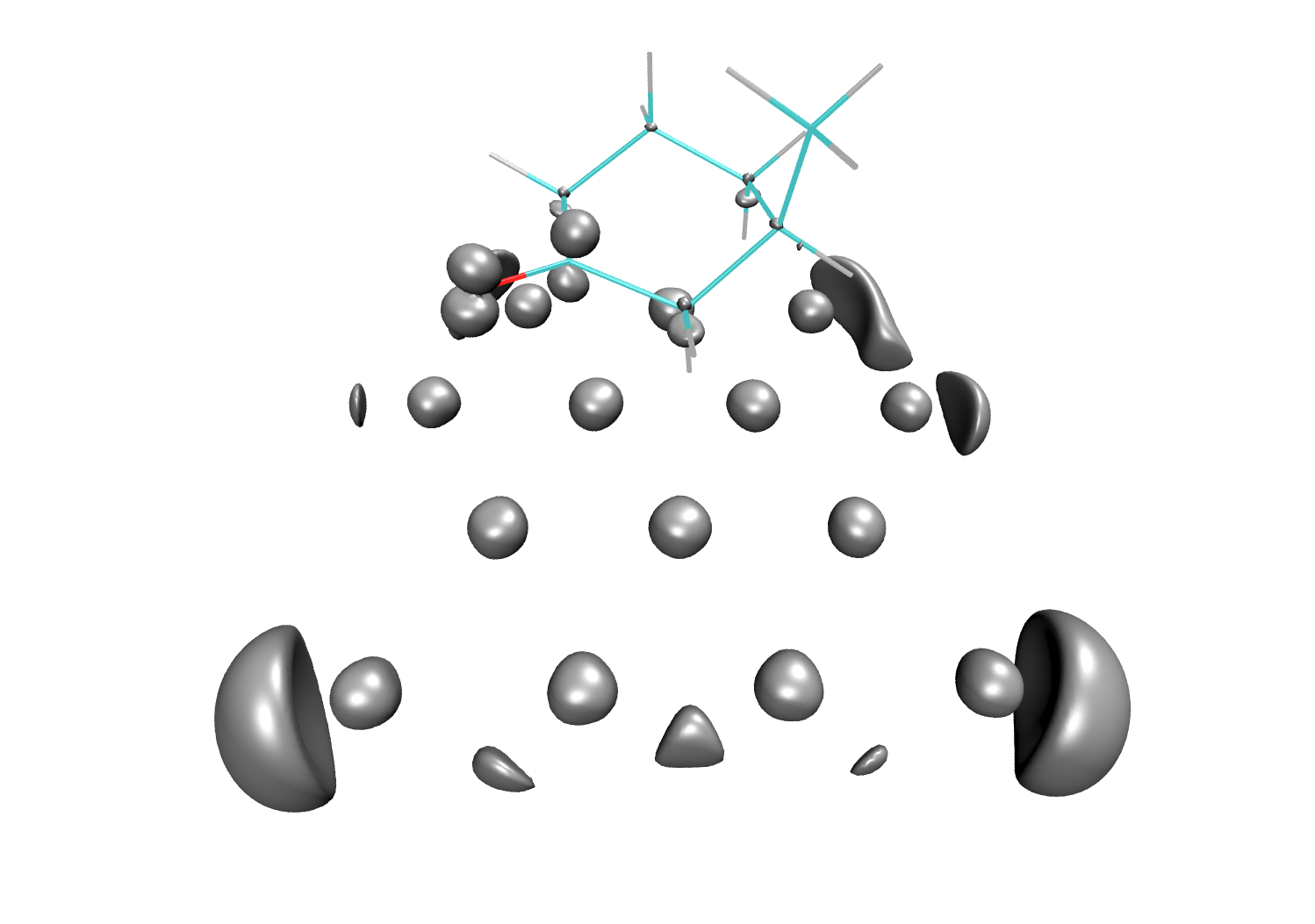}
\includegraphics[clip, trim=3.5cm 3.5cm 3.5cm 3.5cm,width=0.3\linewidth]{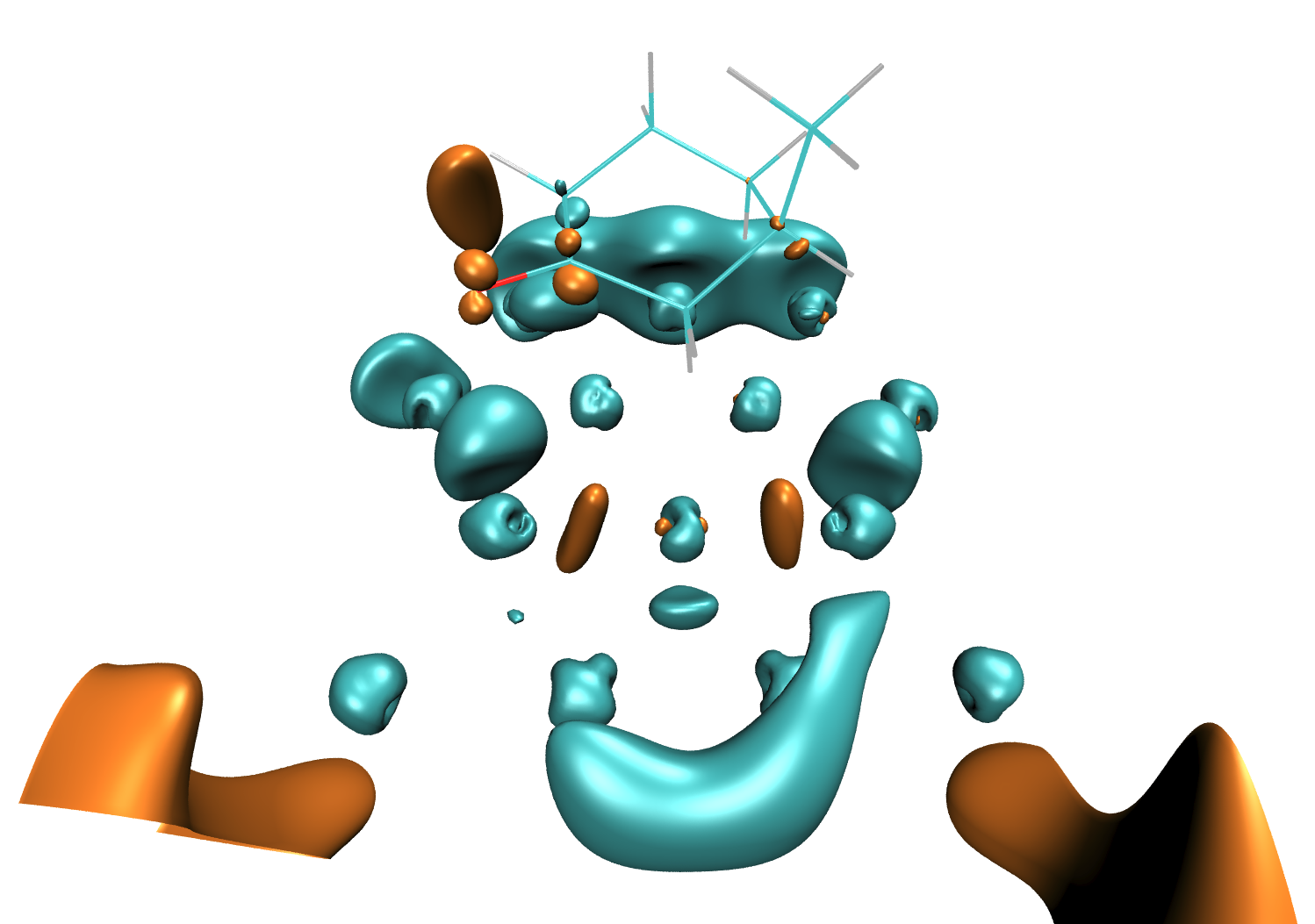}
\caption{\textbf{The photelectron (spin) density of the system.} Left: The system of interest, 3-R-MCHO adsorbed into Cu(111). Middle:  Photoelectron density (equation \ref{eq:nonEqSpinDensityComponent}). Right: Photoelectron spin density resolved along the direction normal to the plane. Cyan/Orange is spin parallel/antiparallel to the surface normal respectively. The surface parallel components are zero, Figure SI-\ref{SI-SI:fig:SpinDensityDiffXYZ}.}
\label{fig:NonEqDensityPIAndPZ}
\end{figure}

\begin{figure}[h!]
\centering
\includegraphics[width=0.55\linewidth]{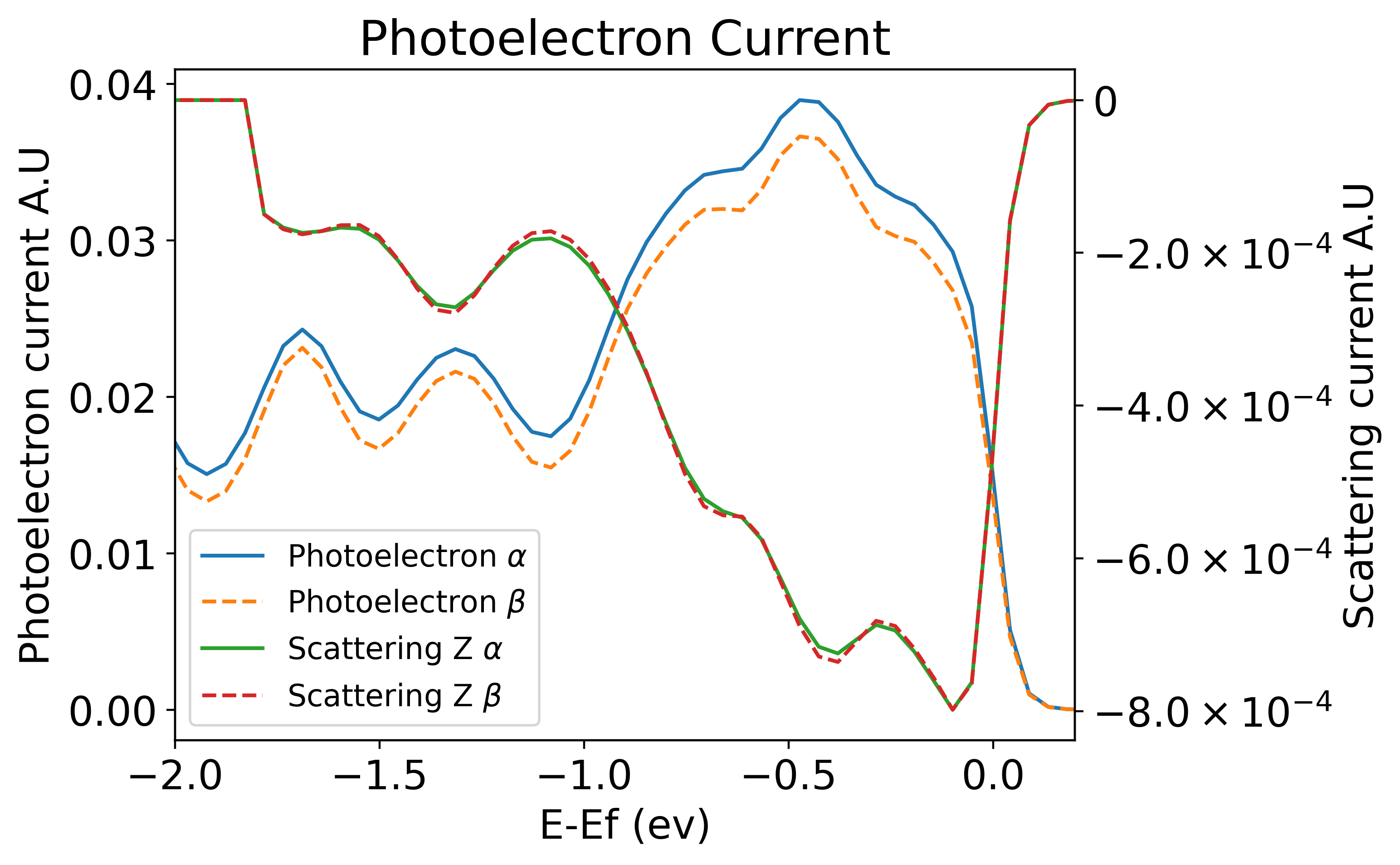}
\includegraphics[width=0.42\linewidth]{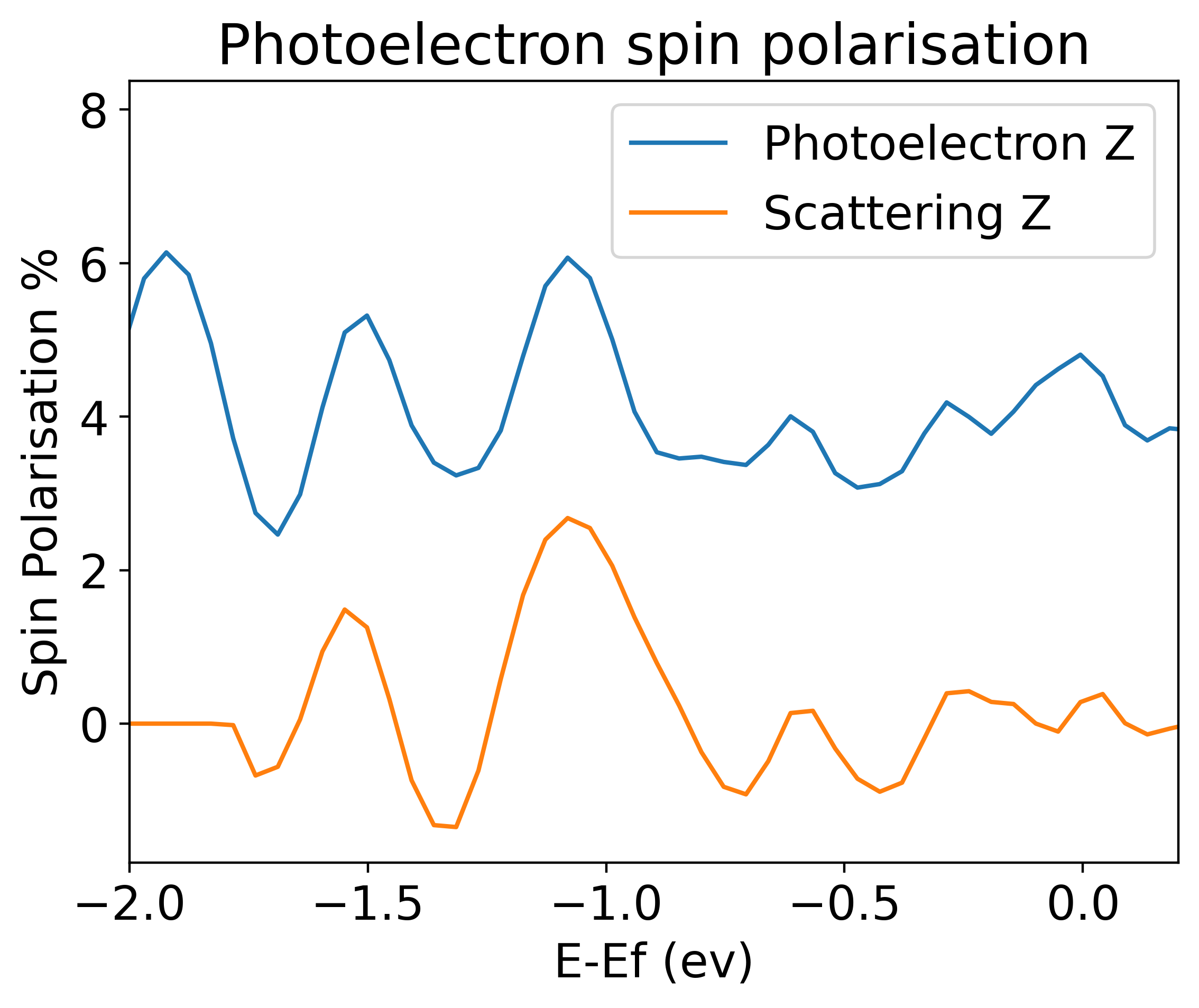}
\caption{\textbf{The magnitude and polarisation of the resultant photoelectrons.} Left: Incoming photoelectron and outgoing (scattering) current. The variation in the current is due to the location of the virtual orbitals of the molecule. Most of the current does not make it across the interface. This is due to the limited delocalisation across the interface, as is typical for physisorbed molecules. The bulk copper was biased by $4\%$ as described in the main text. Right: The spin polarisation of the resultant current. The bias creates an inherent $4\%$ polarisation in the incoming electrons. The outgoing electrons do not have this and all variation is due to the spin-symmetry broken molecular orbitals.   Spin polarisation is as defined by equation \ref{eq:SpinPolEq}. In both cases the spin is resolved along the surface-normal direction.}
\label{fig:SpinPolAndCurrent0.04}
\end{figure}

Figure \ref{fig:NonEqDensityPIAndPZ} shows the distribution of the non-equilibrium density and spin density. The non-equilibrium component of the density matrix is defined by:

\begin{equation}
G^n_{\t{Non-equilibrium Comp.}} = \int_{\mu_{\t{bulk}}}^{\infty} -\iu G^<(E) \td E
\label{eq:nonEqSpinDensityComponent}
\end{equation}

The spin density has a complicated texture. Far away form the chiral molecule the spin texture is approximately symmetric as the surface alone has a $\sigma_v$ mirror plane and so we would expect an approximately symmetric solution. Close to the molecule however, the spin texture is not symmetric due to the influence of the molecules chirality and it is only here that the transport can occur in our calculation. It can be seen that the surface mainly interacts with the CO $\pi$ bond. This suggests that large delocalised structures, such as Helicenes, would exhibit a larger spin textures. There have been multiple reports of CISS with Helicenes\cite{safari_deposition_2022,kettner_chirality-dependent_2018,safari_enantioselective_2024,rodriguez_weakly_2023}\\


Figure \ref{fig:SpinPolAndCurrent0.04} shows the incoming photoelectron and outgoing (scattering) photoelectron current (section SI-\ref{SI-SI:sec:Current}). Most of the electrons do not make it across the interface and are instead scattered back into the copper. This is due to the limited delocalisation of transport channels across the interface. There is an oscillation in both the incoming and scattered current. These oscillations correspond to the changing of the electron transport channels between different MOs. The oscillations are coincident with the fluctuations in the spin polarisation. This is due to each transport channel having its own spin texture. As the electron transport channels change so does the spin of the transport, this leads to oscillations in transmitted spin. This is qualitatively what is seen in the experiment of reference \citenum{badala_viswanatha_vectorial_2022}. It is not expected that the theory here reproduce the experiment perfectly, it is too simple and neglects important features such as the electronic structure of the bulk copper and the interactions with the scattering plane wave states. Nevertheless a surprising agreement in both the magnitude (Expt. $\sim 4\%$) and shape of the photoelectron spectrum is found.\\

 It is stressed that the average $4\%$ polarisation of the incoming photoelectrons (blue line in figure \ref{fig:SpinPolAndCurrent0.04}) is due to the way $\gamma$ was chosen. The CISS signature found is the \textit{oscillations} in the transmitted spin polarisation spectrum, along with the spin texture in figure \ref{fig:NonEqDensityPIAndPZ}. \\

We therefore conclude that the exchange interaction caused by a small ($0.0014\%$) amount of spin buildup greatly influences the resultant photoelectron spectrum leading to a large ($\sim 2\%$) polarisation in transmitted electrons.

\section{Links to Spin Symmetry breaking in Hartree--Fock}

First we must define symmetry breaking. The symmetries we are interested in are those of the Hamiltonian, resp. the experimental setup. If a result of our calculation, say the electron density, does not respect the symmetry of the Hamiltonian it is a symmetry broken solution. At the same time, one can talk about breaking a symmetry of the system. In this case we remove a symmetry of the Hamiltonian by changing the experimental setup. In the above calculation our solution respected the symmetry of the Hamiltonian (SI-\ref{SI-sec:SymmetryConsiderations}).\\

Kohn--Sham (KS) Density Functional Theory (DFT) uses the same single determinant ansatz for its non-interacting wavefunction to generate the density, but has a different effective electron-electron interaction kernel $(h_{pqrs})$.
In both SCF theories (HF and DFT) the spin structure of the ansatz may be constrained in three different ways:
\begin{enumerate}
\item\textbf{R(estricted)HF or R(estricted)KS}: the MOs of the $\alpha$ and $\beta$ spin electrons are the same resulting in an identical spatial distribution of $\alpha$ and  $\beta$ electrons and so there is no spin texture to the density.
\item\textbf{U(nrestricted)HF or UKS}: the MOs may be different for $\alpha/\beta$ electrons but each MO is either $\alpha$ or $\beta$. 
\item\textbf{G(eneralised)HF or non-co-linear DFT}: each MO is now free to have $\alpha$ and $\beta$ components and a varying spin axis. This is important for relativistic theory where $m_{\t{s}}$ is not a \textit{good} quantum number.
\end{enumerate}

The ansatz is usually chosen such that the solution `best' approximates the true solution of the Schr\"{o}dinger equation. Usually `best' is decided based on the symmetry of the problem. Similar spin textures to that found in figure \ref{fig:NonEqDensityPIAndPZ}  are also found in single determinant SCF calculations where the ansatz is not constrained to preserve the spin symmetries\cite{pillai_effect_2025,burton_parity-time_2019}.  This is not a coincidence as at zero temperature in equilibrium the single determinant HF method and the Green's function effective one particle Hamiltonian method are exactly equivalent. The result here best matches that of a UHF determinant. \\

The spin texture found when performing a calculation with a less constrained ansatz is a symmetry-broken solution. It does not respect the symmetry of the Hamiltonian. These less constrained ansatz are used as they can give a better approximation of the ground state energy. This conundrum has been termed L\"{o}wdin's \textit{symmetry dilemma}\cite{lykos_discussion_1963}. Considerable work has gone into understanding why it comes about. It is due to the strength of exchange in HF, that which elsewhere, in a spin symmetric Hamiltonian, creates a symmetry broken solution, creates here an amplified spin texture. Therefore electron-electron exchange is the interaction responsible for the magnitude of the CISS effect.\\

\section{Conclusion}

Until now most theories of CISS have relied predominantly on single non-interacting particle pictures. As is well known to chemists, electrons in molecules cannot be modelled as non-interacting particles and due to interactions complex behaviours can emerge. It is demonstrated here that CISS is such an emergent behaviour. While the origin of CISS likely involves relativistic effects that couple spin to electronic and nuclear motion (spin orbit coupling), the magnitude of the effect can be largely explained by exchange interactions. These exchange interactions can be understood via spin symmetry breaking in Hartree--Fock solutions. These symmetry broken solutions are only realistic if the experiment has already broken the relevant (point group and time-reversal) symmetries. The calculations here show that this can qualitatively reproduce experimental results with the correct magnitude and shape.  This is a promising theory as it does not rely on unphysical parameters, nor on new physics.  All instances of symmetry `breaking' are physically motivated, and the computational approach does not require expensive calculations. The mechanism can be understood using existing literature on spin symmetry breaking in Hartree--Fock theory, allowing a high-level conceptual understanding without complicated mathematics. \\

The current model is simplistic relying on a mean-field approximation and therefore neglecting electron correlations which could influence spin textures. It also relies on the wide-band limit and therefore neglects electronic effects in the bulk copper as well. It also neglects the explicit source of the small initial spin polarisation. Further work is to improve the model via explicit coupling\cite{ozaki_efficient_2010} to bulk copper instead of using the wide-band limit, as well as explicitly coupling to the free photoelectron plane wave states. These would lead to more accurate transport calculations. Another direction is to extend this to larger molecules such as Helicenes and DNA where CISS has been observed experimentally in the tens of percent range. SOC could also be included to remove the artificial spin bias. While explicit coupling to the bulk copper has been developed for DFT, spin symmetry breaking in DFT is less well understood. This is because spin itself is less well quantified in DFT due to Kohn-Sham orbitals not having the same interpretation as HF orbitals.

\section{Acknowledgements}
We would like to express our gratitude to Yuthika Pillai and Andreea Filip for helpful discussions during this work. This work was funded by the Leverhulme Centre for Life in the Universe. 

\section{Author Contributions}
B.C. performed the simulations and developed the code. B.C. and A.J.W.T. analysed and discussed the results. A.J.W.T. supervised the project. All authors contributed to the writing of the manuscript and approved the final version.
\section{Code Availability}
The code used to run these calculations and generate figures is available at \url{https://github.com/apu727/CISSViaExchange}.

\bibliographystyle{unsrt}
\bibliography{CISS}
\end{document}


\title{Supplementary Information for `Explaining the magnitude of Chirality-Induced Spin Selectivity via electron-electron exchange'}
\author{Bence Csakany}
 \email{bc528@cam.ac.uk}
\author{Alex J. W. Thom}
\email{ajwt3@cam.ac.uk}
 \affiliation{Yusuf Hamied Department of Chemistry, University of Cambridge, Lensfield Road, Cambridge, UK.  CB1 2EW}
\maketitle
\section{Symmetry considerations}
\label{sec:SymmetryConsiderations}

Spin symmetry broken Hartree--Fock solutions are usually considered to be an artefact of the approximations made. This arguments rests on the true eigenfunction of the Hamiltonian retaining spin-symmetry. Even with spin-orbit coupling, time-reversal symmetry and point group symmetry can show that the resultant wavefunction has some spin-symmetry and does not exhibit a CISS response. Conversely, a given spin-symmetry broken Hartree--Fock solution has a degenerate pair that is related by the symmetries of the Hamiltonian. Therefore in the case where these symmetries exist, a  spin-symmetry broken Hartree--Fock solution is said to be \textit{unphysical}. In the following we investigate when a CISS response is symmetry allowed, and how this relates to possible Hartree--Fock solutions.\\

We assume the Hamiltonian to have some form of coupling between the spin degrees of freedom and the spatial degrees of freedom. The exact form of this interaction is irrelevant, it suffices that 
\begin{equation}
\left[ \t{e}^{-\iu \v{\sigma}\cdot \v{n} } ,H \right] \neq 0
\end{equation}
Here $\v{\sigma}$ is a vector of Pauli matrices. $\v{n}$ is a vector whose direction indicates the axis of spin-rotation, and whose magnitude indicates the angle by which to rotate. This relation is true for any form of spin-space coupling. \\

We investigate the spin-polarised density CISS signature. The current as discussed in section \ref{SI:sec:Current} is the time derivative of the density in a certain region corresponding to leads.  The spin polarised density is defined along the $x$ or $y$ or $z$ directions as:

\begin{equation}
N_{x,y,z} = \trace{\sigma_{x,y,z} \hat{N} \rho }
\end{equation}
Where $\hat{N}$ is the spatial total particle number operator. $\rho$ is the one particle density matrix. Other directions can be made via linear combinations. For $\sigma_z$ this is:
\begin{equation}
N_{z} =  \rho^{\uparrow\uparrow} - \rho^{\downarrow\downarrow} 
\end{equation}

$ \rho^{\uparrow\uparrow} $ is the up spin, up spin component of the spatial particle number matrix. The total number of electrons is $\rho^{\uparrow\uparrow} + \rho^{\downarrow\downarrow} $. Effectively the one particle density matrix has been split into $4$ spin blocks and traced over the spatial degrees of freedom.

\subsection{Point group Symmetry}
There are an infinite number of point groups. For simplicity we restrict our view to those most typically encountered in a CISS experiment although the argument is general. In a CISS experiment there is generally a molecule adsorbed on a surface. Once adsorbed to the surface the most point group symmetry one can realistically expect in the Hamiltonian is $C_{2v}$. Under this point group it can be shown \cite{altmann_rotations_1986,altmann_point-group_1994} that a single two component spinor $a \ket{\uparrow} + b\ket{\downarrow}$ transforms according to the matrices shown in Table \ref{table:RepsofC2v}. 

\begin{table}[h]
\centering
\begin{tabular}{c|ccccc}
\hline
 & $E$ & $C_2^+$ & $C_2^-$ & $\sigma_{xz}$ & $\sigma_{yz}$  \\
\hline
Trace 
& $2$ & $0$ & $0$ & $0$ & $0$ \\

Rep
& $\begin{pmatrix}1&0\\0&1\end{pmatrix}$
& $\begin{pmatrix}-i&0\\0&i\end{pmatrix}$
& $\begin{pmatrix}i&0\\0&-i\end{pmatrix}$
& $\begin{pmatrix}0&i\\-i&0\end{pmatrix}$
& $\begin{pmatrix}0&-1\\-1&0\end{pmatrix}$\\
$g^{-1} \sigma_z g$ & $\sigma_z$ & $\sigma_z$ & $\sigma_z$ & $-\sigma_z$ & $-\sigma_z$ \\
$g^{-1} \sigma_x g$ & $\sigma_x$ & $-\sigma_x$ & $-\sigma_x$ & $-\sigma_x$ & $\sigma_x$ \\
$g^{-1} \sigma_y g$ & $\sigma_y$ & $-\sigma_y$ & $-\sigma_y$ & $\sigma_y$ & $-\sigma_y$ \\

\hline
\end{tabular}
\caption{Spinor representation of the point group of $C_{2v}$. The rest of the elements of the double group can be found as the negatives of these elements. This representation is in the Cartan gauge. Note that $g^{-1} = g^\dagger$. }
\label{table:RepsofC2v}
\end{table}

The other irreducible representations of $C_{2v}$ can be recovered by taking tensor products of this representation and reducing it. Table \ref{table:RepsofC2v} also shows how each of the Pauli spin matrices transforms under conjugation by the respective point group operator. For a thermal (canonical) ensemble of independent molecules adsorbed on the surface the total Hamiltonian can be written as the direct sum of the individual ones. The density matrix can be written as:

\begin{equation}
\rho = \t{e}^{-\beta H}
\end{equation}
Where $\beta$ is the inverse temperature $1/k_{\t{B}}T$. 
Due to the assumption of individual molecules, the total $H$ retains the same point group symmetries of every molecule. At low densities, molecules adsorbed on a surface can be considered independent. At very high densities, the molecules can have significant interactions with each other, therefore the point group symmetry of the whole surface must be considered instead. 
Clearly the thermal density is unchanged by applying any symmetry operation of the Hamiltonian. We now show that for $C_{2v}$ any CISS signature must vanish in the thermodynamic limit. To see consider the application of the $xz$ mirror plane $\sigma_{xz}$:

\begin{align}
N_{z} = \trace{\sigma_{z} \rho_s } = \trace{(\sigma_{xz} \sigma_{z} \sigma_{xz}) (\sigma_{xz} \rho_s \sigma_{xz})} = -\trace{\sigma_{z} \rho_s }\\
\implies \trace{\sigma_{z} \rho_s } = 0
\end{align}
The other spin components can be found to be zero similarly. Symmetry elements can be removed by changing the molecule adsorbed to the surface. Even if the mirror symmetry is removed (e.g. in a \textit{chiral} molecule) then the $C_{2}$ axis can still be kept (e.g. a helix). In this case we can conclude:

\begin{align}
&N_{z} \neq 0\\
&N_{y} = 0\\
&N_{x} = 0
\end{align}
If the symmetry is broken such that only the $xz$ mirror plane, $\sigma_{xz}$, is kept then:
\begin{align}
&N_{z} = 0\\
&N_{y} \neq 0\\
&N_{x} = 0
\end{align}
with an analogous result if only $\sigma_{yz}$ is kept. This result shows that in a achiral molecule, point-group symmetry allows a spin polarisation of the density only in certain directions. 

\subsection{Time reversal symmetry}
Time-reversal is an anti-unitary operator defined by the operation:
\begin{equation}
\mathcal{T} = -\iu \sigma_y \mathcal{K} \t{ and } t \rightarrow -t
\end{equation}
Where $\mathcal{K}$ is the complex conjugation operator that complex conjugates everything to the right of it. It can be verified that the following identities are true \cite{barron_time_nodate,burton_parity-time_2019} for two time independent wavefunctions $\ket{\psi}$ and $\ket{\phi}$:

\begin{align}
\braket{\mathcal{T}\psi|\mathcal{T}\phi} &= \braket{\psi^*|\phi^*} = \braket{\phi|\psi}\label{eq:TimeReversalAntiUnitary}\\
\mathcal{T}\ket{\phi}\bra{\psi}(-\mathcal{T}) &= \ket{\psi}\bra{\phi}\\
\mathcal{T}^{-1} = \iu \sigma_y \mathcal{K} &= -\mathcal{T}\\
\mathcal{T}\v{\sigma} = -\v{\sigma} \mathcal{T} \label{eq:SigmaUnderT}
\end{align}
Where $\v{\sigma}$ is the vector $\vector{\sigma_x,\sigma_y,\sigma_z}$.\\

We can therefore make the same argument as before and consider
\begin{align}
N_{z} = \trace{\sigma_{z} \rho_s } = \trace{(-\mathcal{T})\sigma_{z} \mathcal{T} (-\mathcal{T})\rho_s\mathcal{T}  } = -\trace{\sigma_{z} \rho_s}\\
\implies \trace{\sigma_{z} \rho_s } = 0
\end{align}
likewise for all other spin components. Therefore if $\mathcal{T}$ is a symmetry of the Hamiltonian, there can be no spin-polarisation in any direction. Although not relevant in the photoelectron case, in general one should consider if the combined action of $\mathcal{T}$ with a spatial symmetry e.g. $C_{2}\mathcal{T}$ is a symmetry of the Hamiltonian even if $\mathcal{T}$ and $C_2$ are not individually symmetries. As an example this applies if the molecule has a spatial symmetry that is broken by an external magnetic field\cite{barron_true_1986}.

\subsection{Wavefunctions} 
\label{SI:Wavefunctions}
The above arguments are valid for a thermal ensemble of molecules. This is valid if the molecules adsorbed to the surface can be considered to be independent of each other. Each individual molecule can still support a spin polarisation even with time-reversal and point group symmetry only if the ground-state energy level is degenerate. It is their thermal average that has vanishing spin polarisation. \\

To see this we repeat the arguments above but for a single coherent state $\ket{\psi}$, taken to be the ground state of the Hamiltonian. Since $g$ is a symmetry of the Hamiltonian, $\ket{\psi}$ and $g\ket{\psi}$ are both eigenfunctions of the Hamiltonian.

\begin{align}
H\ket{\psi} = E\ket{\psi}\\
H g \ket{\psi} = g H \ket{\psi} = E g \ket{\psi}
\end{align}

For time-reversal the situation is slightly changed as it is an anti-unitary operator. The Hamiltonian of a closed quantum system must always be Hermitian such as to preserve the interpretation of $|\psi |^2$ as a probability distribution. Approximate Hamiltonians can be non-Hermitian, however, if all interactions of relevance are included, the Hamiltonian must be Hermitian. This means that the `true' $H$ always has real eigenvalues. Considering the argument above for time reversal therefore gives:
\begin{align}
H\ket{\psi} = E\ket{\psi}\\
H \mathcal{T} \ket{\psi} = \mathcal{T} H \ket{\psi} =  E^* \mathcal{T}\ket{\psi} = E \mathcal{T}\ket{\psi}
\end{align}
Once again both $\ket{\psi}$ and $\mathcal{T}\ket{\psi}$ are eigenfunctions of the Hamiltonian. If the eigenstate is non-degenerate we require that 
\begin{equation}
g\ket{\psi} = \lambda \ket{\psi} \label{eq:pointGroupEigenFunction}
\end{equation}
For point group symmetry this is an eigenfunction equation. The time-reversal operator does not have eigenfunctions and so a unique $\lambda$ cannot be assigned to each wavefunction. Nevertheless since both $\ket{\psi}$ and $\mathcal{T}\ket{\psi}$ are non-degenerate eigenfunctions of $H$ that also preserve $\braket{\psi|\psi}$. Therefore $\ket{\psi}$ and $\mathcal{T}\ket{\psi}$ must differ by at most a unit modulus phase factor. \\

In the non-degenerate case we can make a similar argument to before for some point-group symmetry (Unitary) operation satisfying $g^{-1} \sigma_z g = -\sigma_z$:

\begin{align}
\braket{\psi |\sigma_z|\psi} = \braket{g \psi |\sigma_z| g \psi} = \braket{\psi | g^\dagger \sigma_z g | \psi} = -\braket{\psi |\sigma_z|\psi}
\end{align}
and likewise for the other spin components. For time-reversal, anti-unitary, the argument is:
\begin{align}
\braket{\psi |\sigma_z|\psi} \overset{\t{using eq. \ref{eq:TimeReversalAntiUnitary} }}{=} \braket{T \psi |\sigma_z| T \psi} = \braket{\psi | (-T) \sigma_z T | \psi}^* = -\braket{\psi |\sigma_z|\psi}
\end{align}
where we have used that $\braket{\psi |\sigma_z|\psi}$ is a real number since $\sigma_z$ is Hermitian, along with:
\begin{align}
\braket{\mathcal{T}\psi |\sigma_z|\mathcal{T}\psi}=&\int (-\iu \sigma_y \mathcal{K} \psi)^* \sigma_z (-\iu \sigma_y \mathcal{K} \psi) \td \tau\\
& \int \psi (+\iu \sigma_y) \sigma_z (-\iu \sigma_y) \psi^* \td \tau = \\
& \left(\int \psi^* (+\iu \sigma_y\mathcal{K}) \sigma_z (-\iu \sigma_y\mathcal{K}) \psi \td \tau\right)^* \\
&= \braket{\psi | (-\mathcal{T})\sigma_z\mathcal{T}|\psi}^*
\end{align}

In the degenerate case we can no longer conclude that:
\begin{equation}
g\ket{\psi} = \lambda \ket{\psi} \tag{\ref{eq:pointGroupEigenFunction}}
\end{equation}
Defining $\ket{\phi} = g\ket{\psi}$, we can however conclude that $\ket{\phi}$ is still an eigenfunction of $H$, but it may be a different one. In this case:

\begin{equation}
\braket{\psi |\sigma_z|\psi} = -\braket{\phi |\sigma_z|\phi}
\end{equation}
i.e. $\ket{\psi}$ and $\ket{\phi}$ are two degenerate wavefunctions with opposite CISS signatures. The previous argument based on the density can be recovered by forming the incoherent superposition of these two eigenstates:

\begin{equation}
\rho = \ket{\psi}\bra{\psi} + \ket{\phi}\bra{\phi}
\end{equation}
which has no CISS signature as can be verified. The thermal density matrix naturally forms these superpositions. If the molecules cannot be considered independent then this superposition is invalid. A example of a case where the molecules cannot be considered independent is a bar magnet. In a bar magnet each spin centre cannot be considered independent. Both polarisations of a bar magnet are degenerate in energy as expected from symmetry, yet the magnet still has a defined direction. Note that two different bar magnets are independent of one another and therefore bar magnets with both polarisations can be found.

\subsection{Photoelectrons}
\label{SI:photoelectrons}
In a photoelectron experiment, the incoming photon flux breaks time-reversal symmetry. This is evident as reversing the flow of time would reverse the direction of the incoming photons turning a photoelectron experiment into an inverse-photoelectron experiment. Likewise the outgoing photoelectron flux is a true chiral phenomenon \cite{barron_true_1986}. In the experiment considered here, the molecule-surface-incoming photon system has no spatial symmetry. The molecule is chiral -- precluding mirror planes and inversion. The molecule has no rotational symmetries. Therefore neither point-group symmetry nor time-reversal symmetry suggest a zero spin polarisation. Therefore a non spin-symmetric UHF like solution -- as found -- is a \textit{physical} approximation to the true wavefunction/density.

\subsection{Hartree--Fock}

Hartree--Fock determinants are not solutions to the true Schr\"{o}dinger equation, therefore even if the true eigenstate is non-degenerate, there may be multiple solutions to the Hartree--Fock equations. The energy of the Hartree--Fock determinant is however calculated using the true $H$. Applying a symmetry operation $g$ of the true Hamiltonian to a Hartree--fock solution will either yield the same solution, in which case the non-degenerate argument of section \ref{SI:Wavefunctions} follows, or a different solution. If it yields a different solution and the ground state is not degenerate, then the solution is said to be symmetry broken. In this case the degenerate argument of section \ref{SI:Wavefunctions} follows. The two solutions do not accurately represent the true ground states properties as far as CISS signatures are concerned. In this regard Hartree--Fock solutions are \textit{unphysical}. In the calculations performed here, as discussed in section \ref{SI:photoelectrons} there are no such symmetries and therefore the Hartree--Fock solutions are \textit{physical} approximations to the ground state.

\section{Green's functions in a non-orthogonal basis}
\label{SI:NonOrthGreensFuncSec}
Quantum mechanical Green's functions can be formulated\cite{artacho_nonorthogonal_1991} in non-orthogonal basis. By associating annihilation operators as annihilation operators on the direct space and creation operators as annihilation operators on the dual space, both operators can be ascribed upper and lower indices. For a basis function $i$ this corresponds to the modification:

\begin{align}
(a_i)^\dagger &\rightarrow a_i\\
(a_i) &\rightarrow a^i
\end{align}
In this notation a one electron wavefunction is written using summation convention as:
\begin{equation}
\psi = c^i a_i \ket{}
\end{equation}
The new creation--annihilation operators have the same algebra as their orthogonal counterparts. Complex conjugation converts a bra into a ket giving:
\begin{equation}
\psi^* = \bra{} c^{\bar{i}} a_{\bar{i}}
\end{equation}
The bar $\bar i$ above an index denotes that is has been complex conjugated.  
Note how the annihilation operator is no longer the complex conjugate of the creation operator. In a finite basis it is however still the adjoint and can be related using the metric. 

\begin{equation}
\psi = c^i a_i \ket{} = \left(\bra{}c^{\bar{i}} S_{\bar{i}j} a^j \right)^*
\end{equation}

Alternatively and equivalently one may work in a symmetrically orthogonalised basis such as the L\"{o}wdin basis. This is the basis transformation given by:
\begin{align}
&c_i = (S^{1/2})_i^{\ j} (c_{\t{lowdin}})_j\\
&c^i = (S^{-1/2})^i_{\ j} (c_{\t{lowdin}})^j
\end{align}
for some coefficient vector $\v{c}$. The indentation of the indices refers to the rows/columns of the matrix. i.e. $(S^{-1/2})_i^{\ j} \equiv (S^{-1/2})_{ij}$

\section{Self Energy Terms}
\label{SI:SelfEnergyTerms}
The solution to a Green's function problem is given by the solution to the non-equilibrium Dyson equations\cite{datta_electronic_1995,rammer_quantum_2007}. Using summation convention:

\begin{equation}
\begin{split}
\left(E \delta^{i}_{k} - H^{i}_{k} - (\Sigma^R)^{i}_{k}\right)G^R(E)^{k}_{j} = \delta^{i}_{j},\\
\left(E \delta^{i}_{k} - H^{i}_{k} - (\Sigma^A)^{i}_{k}\right)G^A(E)^{k}_{j} = \delta^{i}_{j},\\
\left(E \delta^{i}_{k} - H^{i}_{k} \right)G^K(E)^{k}_{j} = (\Sigma^R)^{i}_{k}G^K(E)^{k}_{j} + (\Sigma^R)^{i}_{k} G^K(E)^{k}_{j}.
\end{split}
\label{eq:DysonMatrixEquations}
\end{equation}

\begin{align}
H &= H^{\t{kin}} + H^{\t{nuc}}\\
H^{\t{kin}}_{\bar{i}j} &= \braket{\phi_i|-\half \nabla^2 | \phi_j} \\
H^{\t{nuc}}_{\bar{i}j} &= \sum_{I} \braket{\phi_i|\frac{-1}{\lvert r-R_I\rvert } | \phi_j} \\
\end{align}
Where $\phi_i$ are atomic orbitals. $R_I$ are the nuclear coordinates. $r$ is the spatial coordinate of the atomic orbitals. 

The self energies are:

\begin{widetext}
\begin{equation}
(\Sigma^{R/A/K})^{i}_{k} = (\Sigma^{R/A/K}_{\t{HF}})^{i}_{k} + (\Sigma^{R/A/K}_{\t{bulk}})^{i}_{k} + (\Sigma^{R/A/K}_{\t{photoelectrons}})^{i}_{k} + (\Sigma^{R/A/K}_{\t{vacuum}})^{i}_{k}
\end{equation}
\end{widetext}
$(\Sigma^{R/A/K}_{\t{HF}})$ is the Hartree--Fock self energy. $(\Sigma^{R/A/K}_{\t{bulk}})$ is the coupling of the system to the bulk copper states. $(\Sigma^{R/A/K}_{\t{photoelectrons}})$ is the coupling to the excited photoelectron states in bulk copper. $(\Sigma^{R/A/K}_{\t{vacuum}})$ is the coupling to the outgoing free electron states in the vaccum. \\

The Hartree--Fock self energy is dependent on the density and is given by:

\begin{align}
(\Sigma^{R}_{\t{HF}})_{\bar{q}s} &= (\Sigma^{A}_{\t{HF}})_{\bar{q}s} = \frac{1}{2\pi}(h_{\bar{p}\bar{q}rs}-h_{\bar{q}\bar{p}rs}) (G^n)^{r\bar{p}}\\
h_{\bar{p}\bar{q}rs} &= \braket{q(\bfr_1)p(\bfr_2)|\frac1{\lvert \bfr_1-\bfr_2\rvert} |r(\bfr_2)s(\bfr_1)}\\
(\Sigma^{K}_{\t{HF}})_{\bar{q}s} &= 0\\
(G^n)^{r\bar{p}} &= \int_{-\infty}^{\infty} -\iu (G^<)^{r\bar{p}}(E) \td E \label{SI:eq:GnDefinition}
\end{align}

If the interaction between the leads and the molecule is a one electron interaction, the exact coupling is given by\cite{datta_electronic_1995}:

\begin{equation}
(\Sigma^{R/A/K}_{\t{bulk}})^{i}_{l} = U^i_j (G^{R/A/K}_{\t{bulk}})^j_k U^k_l 
\end{equation}
However this requires knowledge of the true Greens function for bulk copper. The density of states of the bulk copper is related to $G^R$ via:

\begin{equation}
A = \iu (G^R - G^A)
\end{equation}
where $A$ is the spectral density of states. Thus it is common to approximate $(G^{R}_{\t{bulk}})$ as a pure constant negative imaginary number corresponding to $\half$ the density of states. This is called the wide-band limit. Specifically here the self energy for the bulk copper interacting with the copper surface is:

\begin{widetext}
\begin{equation}
(\Sigma^{R}_{\t{bulk}})_{ij} = 
\begin{cases}
\frac{-\iu \gamma^{\uparrow}_{b}}{2} & \t{if } i = j \t{ and } i \t{ is on the copper surface} \t{ and } i \t{ is spin up} \\
\frac{-\iu \gamma^{\downarrow}_{b}}{2} & \t{if } i = j \t{ and } i \t{ is on the copper surface} \t{ and } i \t{ is spin down} \\
0 & \t{otherwise} 
\end{cases}
\end{equation}
\end{widetext}
This tensor is in the written in the L\"{o}wdin basis (section \ref{SI:NonOrthGreensFuncSec}) as these approximately correspond to localised orbitals. Therefore in the atomic orbital basis it is:

\begin{equation}
(\Sigma^{R}_{\t{bulk}})^{i}_{l} = (S^{-1/2})^{i}_{\ j} (\Sigma^{R}_{\t{Bulk  copper}})_{jk} (S^{1/2})_{k}^{\ l}
\end{equation}
the parameters $\gamma^{\uparrow}_{b}$ and $\gamma^{\downarrow}_{b}$ represent the combined density of states and interaction energy of the bulk copper. They are split along the spin directions in order to provide an initial biased spin density as discussed in the main text. $(\Sigma^{K}_{\t{bulk}})^{i}_{l}$ is given by:
\begin{equation}
(\Sigma^{K}_{\t{bulk}})^{i}_{l} = (1-2f(E;\mu_b))((\Sigma^{R}_{\t{bulk}} - \Sigma^{A}_{\t{bulk}})^{i}_{l})
\end{equation}
This assumes the bulk copper is in thermal equilibrium with chemical potential $\mu_b$. $f(E;\mu_b)$ is the Fermi-Dirac distribution at energy $E$ and chemical potential $\mu_b$. The photoelectron coupling term $(\Sigma^{R/A/K}_{\t{photoelectrons}})^{i}_{k}$ has the same form but with a different $\mu_p$, $\gamma^{\uparrow}_p$ and $\gamma^{\downarrow}_p $.
\\ 

Similarly the coupling to the scattering states is given by:
\begin{widetext}
\begin{equation}
(\Sigma^{R}_{\t{vacuum}})_{ij} = 
\begin{cases}
\frac{-\iu \gamma^{\uparrow}_{s}}{2} & \t{if } i = j \t{ and } i \t{ is on 3-MCHO} \t{ and } i \t{ is spin up} \\
\frac{-\iu \gamma^{\downarrow}_{s}}{2} & \t{if } i = j \t{ and } i \t{ is on 3-MCHO} \t{ and } i \t{ is spin down} \\
0 & \t{else} 
\end{cases}
\end{equation}
\end{widetext}
for $E > 0$ and is $0$ otherwise. This represents the ability for the vaccum to accept states only if they have enough energy to escape. The `chemical potential' of the vaccum is set below $0$ such that it is always empty of electrons. The parameters used are summarised in table \ref{table:Parameters_used}. \\

The choice of $\gamma_{\t{Bulk Copper}} = 10^{-2}\Eh$ can be obtained from the density of states of bulk copper. A reasonable number for the density of states at the Fermi level of copper is $0.3\t{eV}^{-1}\t{atom}^{-1} = 8\t{E}_{\t{h}}^{-1}\t{atom}^{-1}$. If the interaction term between neighbouring copper atomic orbitals in the band of interest is on the order of $0.1\t{E}_\t{h}$ as is typical, this gives:
\begin{equation}
\gamma = 2\pi U^2 \t{D.O.S} = 2\pi\times8\times 10^{-2}\t{E}_{\t{h}} \approx 48 \times 10^{-2}\t{E}_{\t{h}}
\end{equation}
Therefore $\gamma = 10^{-2}\Eh$ underestimates the true parameter. Approximating $\gamma$ in this way is crude as different bands have different interaction energies and different density of states. Further the density of states is different in the bulk to at the surface. Increasing it would induce a larger equilibrium spin polarisation and increase the CISS effect seen. \\

\begin{table}[ht!]
\centering
\begin{tabular}{c|c c}
Quantity & Value & Unit\\
\hline
$\gamma_b^{\uparrow} $ & $1.04 \times 10^{-2}$ & $\Eh$\\
$\gamma_b^{\downarrow} $ & $0.96 \times 10^{-2}$ & $\Eh$\\
$\gamma_s^{\uparrow} $ &  $10^{-2}$ & $\Eh$\\
$\gamma_s^{\downarrow} $ &$10^{-2}$ & $\Eh$\\
$\gamma_p^{\uparrow} $ & $1.04 \times 10^{-4}$ & $\Eh$\\
$\gamma_p^{\downarrow} $ & $0.96 \times 10^{-4}$ & $\Eh$\\
$\mu_b $ & $-0.15$ & $\Eh$\\
$\mu_p $ & $-0.15\Eh+5.9\t{eV}$ & \\
$T$ & $300$ & K

\end{tabular}
\caption{Parameters used to setup the calculation unless otherwise stated. $\gamma$ terms are the coupling strength to the 3 different leads. $\mu$ are the chemical potentials of the different leads. Parameters are described in more detail in section \ref{SI:SelfEnergyTerms}}
\label{table:Parameters_used}
\end{table}

 Since $(\Sigma^{R}_{\t{HF}})$ depends on the density. The Dyson equations are solved iteratively, updating the density with the newest solution until convergence is reached. Direct inversion in the iterative subspace (DIIS) techniques \cite{pokhilko_iterative_2022} accelerate the convergence. 
\section{Calculation Specifics} 
The atomic orbital basis used was the LANL2DZ basis. The geometry of the adsorbed 3-MCHO was optimised using the PBE functional in QCHEM while keeping the geometry of the copper layer fixed with a bond length of $2.55\t{\AA}$. Correctness of the implementation was verified by comparing the equilibrium RHF energies (Custom code: $-3076.958\Eh$, QCHEM: $3076.941\Eh$). The disagreement in the $6^\t{th}$ significant figure is likely due to integral generation specifics and has no effect on the result here. Calculations were converged to a DIIS error of $5\times 10^{-10}\Eh$. All input files, geometries and code are available at \url{https://github.com/apu727/CISSViaExchange}.

\section{Expression for the current}
\label{SI:sec:Current}
It can be shown \cite{rammer_quantum_2007} that the current entering/leaving the system due to the interactions with lead $p$ is:
\begin{align}
I_p(E) \equiv \DbyD{t} N(E) = \frac{-1}{4\pi} \trace{ \Sigma^R_p G^K + \Sigma^K_p G^A - G^K \Sigma^A_p - G^R \Sigma^K_p}
\end{align}
A positive $I_L$ means conventional current is flowing out of lead $L$ and into the system. As this is the derivative of a physical quantity, it is gauge invariant. $N(E)$ is the total number of electrons in the system at a given single particle energy. The total number of electrons can be resolved along spin directions as in section \ref{sec:SymmetryConsiderations} to give a spin-resolved current. It is equivalent to the expression for current in the Landauer-B\"{u}ttiker formalism:

\begin{align}
I_p(E) &= \sum_q -T_{pq} f(E_q;\mu_q) + \sum_{q} T_{qp} f(E_p;\mu_p)\\
T_{pq} &= \frac{1}{2\pi} \trace{\Gamma_p G^R \Gamma_q G^A}\\
\Gamma_p &= \iu \left(\Sigma^R_p - \Sigma^A_p\right)
\end{align}

Or alternatively a third form commonly seen in literature\cite{datta_electronic_1995}:

\begin{align}
I_p(E) = \frac{1}{2\pi} \trace{\Sigma^<_p G^>_p - \Sigma^>_p G^<_p }
\end{align}

In all cases this gives the energy resolved current and can be integrated to give the total current. It can also be split along spin boundaries to give a spin resolved current. 

\section{Supplementary Figures}
\label{SI:sec:DifferentParameters}

\begin{figure}[ht!]
    \centering
    \includegraphics[width=0.5\linewidth]{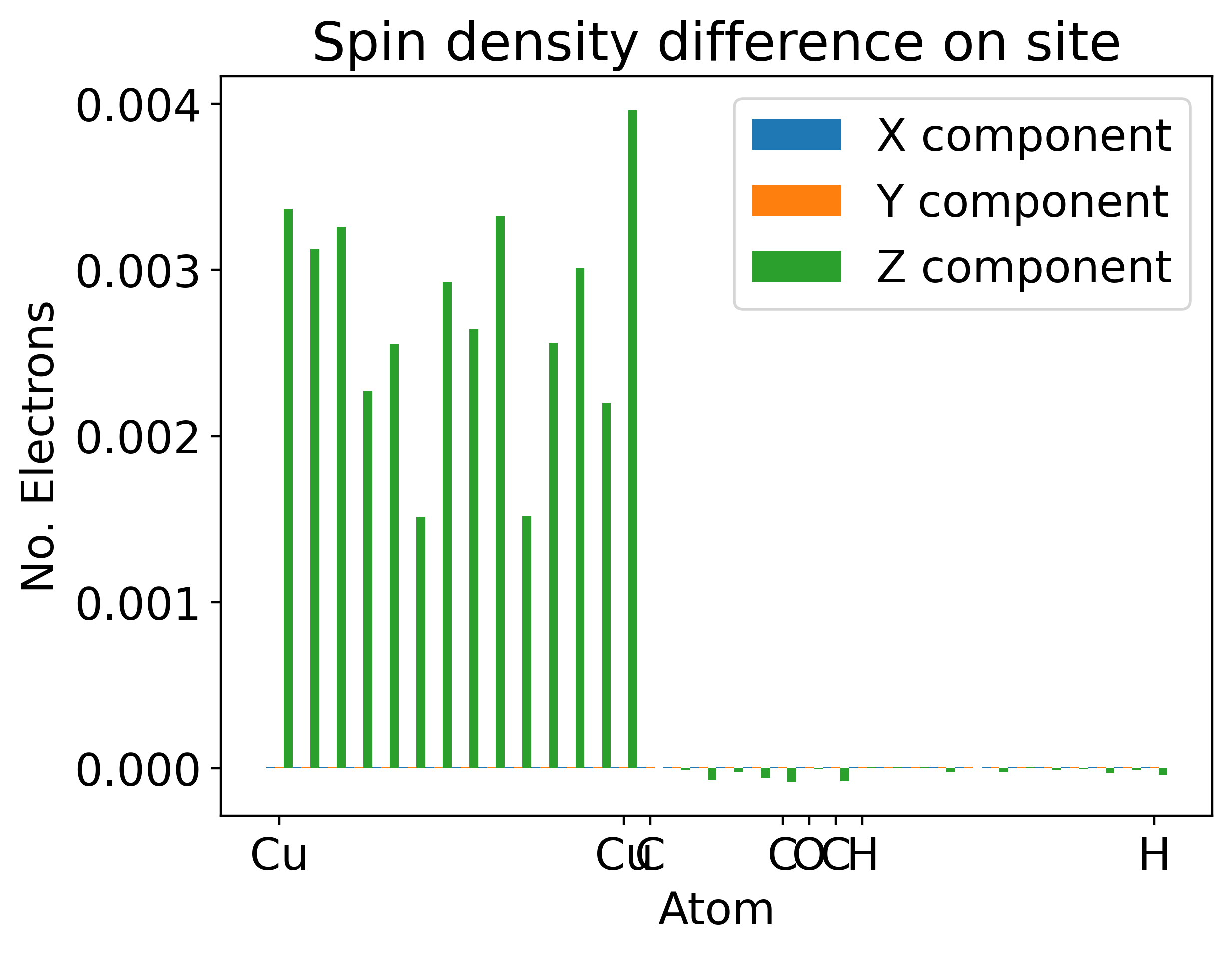}
    \caption{Spin difference in number of electrons (equation \ref{SI:eq:SpinDensityDiff}). The $X$, $Y$ component of the spin density difference is zero. This is because spin-orbit coupling was not included in the calculation, therefore each spin `chooses' to align parallel, or anti-parallel. Parameters are as shown in table \ref{table:Parameters_used}. This is using the full $G^n$ not the non-equilibrium component, equation Main Text-\ref{MT-eq:nonEqSpinDensityComponent}. } 
    \label{SI:fig:SpinDensityDiffXYZ}
\end{figure}
Figure \ref{SI:fig:SpinDensityDiffXYZ} shows the spin density difference on each site resolved in the $X$,$Y$,$Z$ directions. The $X$,$Y$ directions lie in the plane of the surface. The $Z$ direction is out of the plane. The on site spin density difference is found using the non equilibrium Green's function $G^n$ (equation \ref{SI:eq:GnDefinition}) .

\begin{equation}
n_{I,s} = \frac{1}{2\pi} \sum_{i \in I} \sum_{\mu\nu} \sigma_{\mu\nu} (G^n)^{i\nu}_{i\mu} \label{SI:eq:SpinDensityDiff}
\end{equation}
Where $s$ is either $X$,$Y$ or $Z$. $i$ denotes an atomic orbital. $I$ denotes the atom. $i \in I$ denotes all the atomic orbitals on a given atom. The double index $i\mu$ denotes an atomic orbital $i$ with spin $\mu$. The spin polarised current is defined as:
\begin{equation}
\t{Spin polarised current} = \frac{I_{p\uparrow} - I_{p\downarrow}}{I_{p\uparrow} + I_{p\downarrow}} \times 100\%
\end{equation}

\begin{figure}[ht!]
    \centering
    \includegraphics[width=0.45\linewidth]{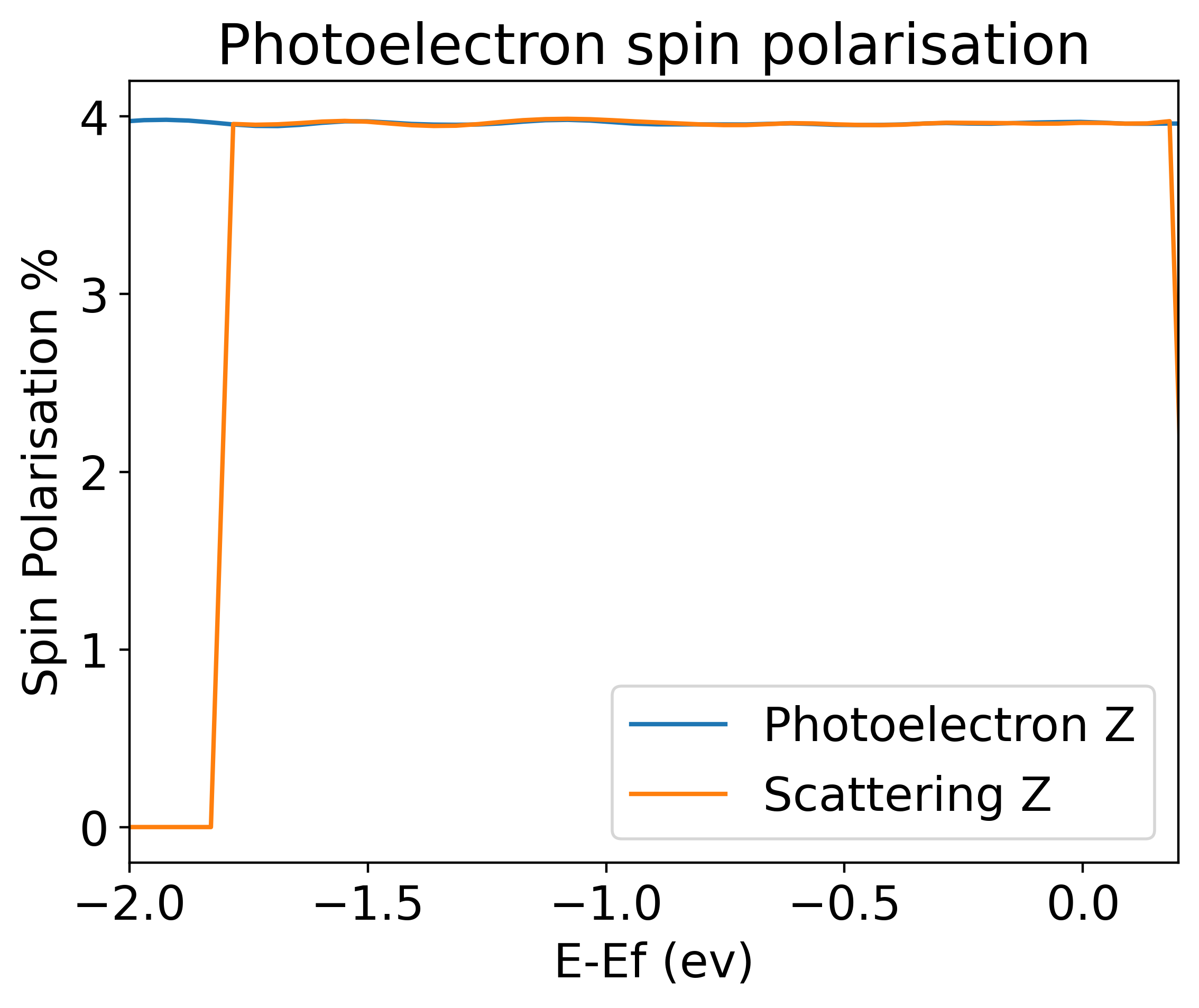}
    \includegraphics[width=0.45\linewidth]{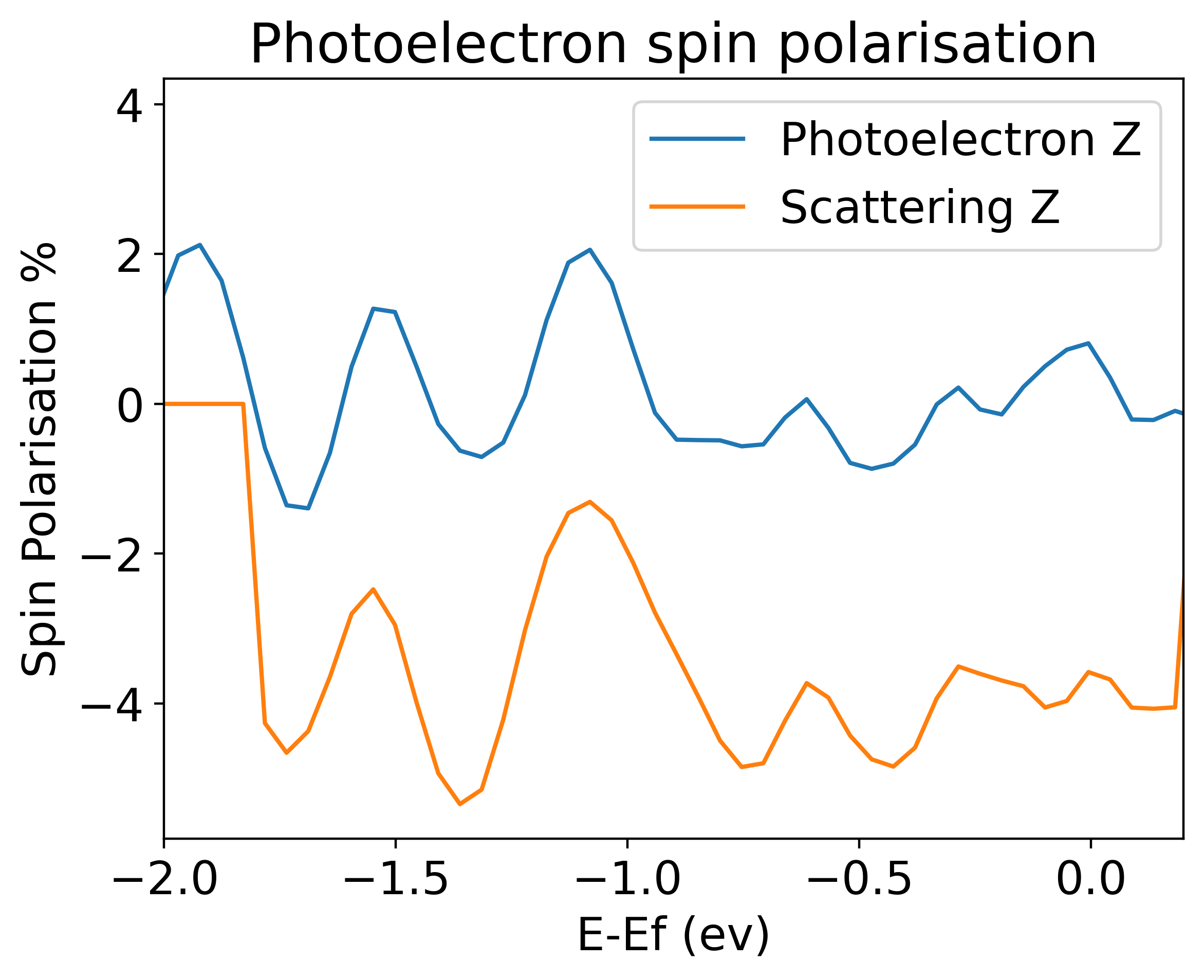}
    \caption{Energy resolved spin polarised current along $Z$ direction. Left: the parameters were modified such that there is no spin bias from the bulk copper, only the photoelectrons. Since this leads to a much smaller spin-symmetry breaking the CISS effect is much smaller. Right: No spin polarisation of photoelectrons, but spin polarisation of bulk copper. Parameters are as shown in table \ref{table:Parameters_used} except for. Left: $\gamma_b^{\uparrow} = \gamma_b^{\downarrow}  = 10^{-2}\Eh$. Right: $\gamma_p^{\uparrow} = \gamma_p^{\downarrow}  = 10^{-2}\Eh$}
    \label{SI:fig:SpinPolarisedCurrentwithNoSpinBias}
\end{figure}

Figure \ref{SI:fig:SpinPolarisedCurrentwithNoSpinBias} left shows the photoelectron current spin polarisation if there is no bulk copper spin symmetry breaking. The $4\%$ polarisation is an artefact of the calculation. This is due to the incoming photoelectrons $\gamma_s$ having a $4\%$ polarisation in $\gamma_p$ but not in $\gamma_b$. This $4\%$ polarisation then directly transfers to the two outgoing electron streams, back scattering into the copper, and forward scattering into the free particle scattering states. The CISS signature is the variation in the polarisation of the resultant electrons which is much lower. In contrast when $\gamma_b$ is spin polarised but $\gamma_p$ is not, figure \ref{SI:fig:SpinPolarisedCurrentwithNoSpinBias} right, the CISS signature can clearly be seen due to the spin-symmetry broken solution. In both cases the scattering current drops off at $E-E_f = -1.8\t{eV}$ which is when the energy of the photoelectrons is below $0 \t{eV}$. Note that the photoelectron current does not drop as scattering back into the copper surface is still allowed.

\subsection{Stability of effect with variation of parameters}

\begin{figure}[ht!]
    \centering
    \includegraphics[width=0.45\linewidth]{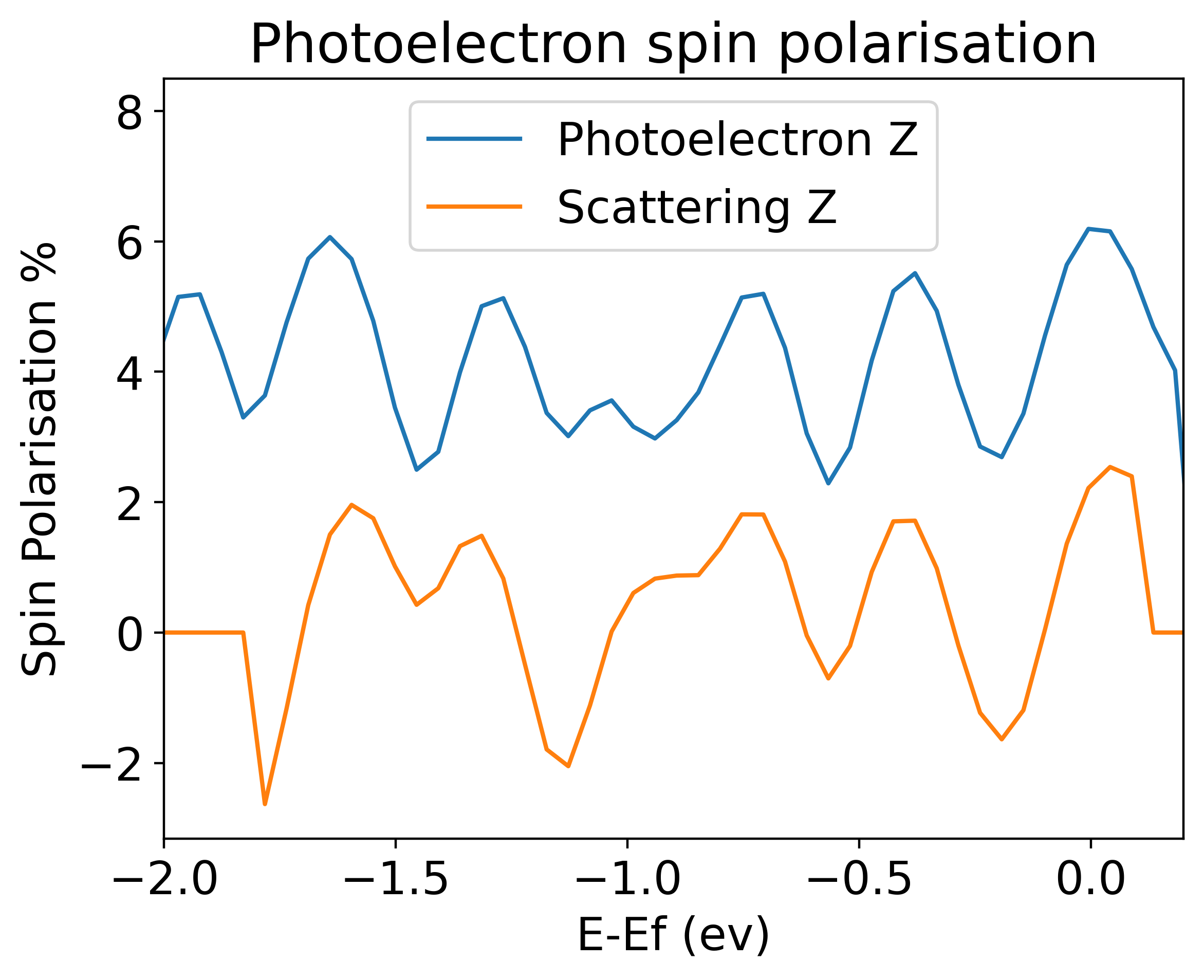}
    \includegraphics[width=0.45\linewidth]{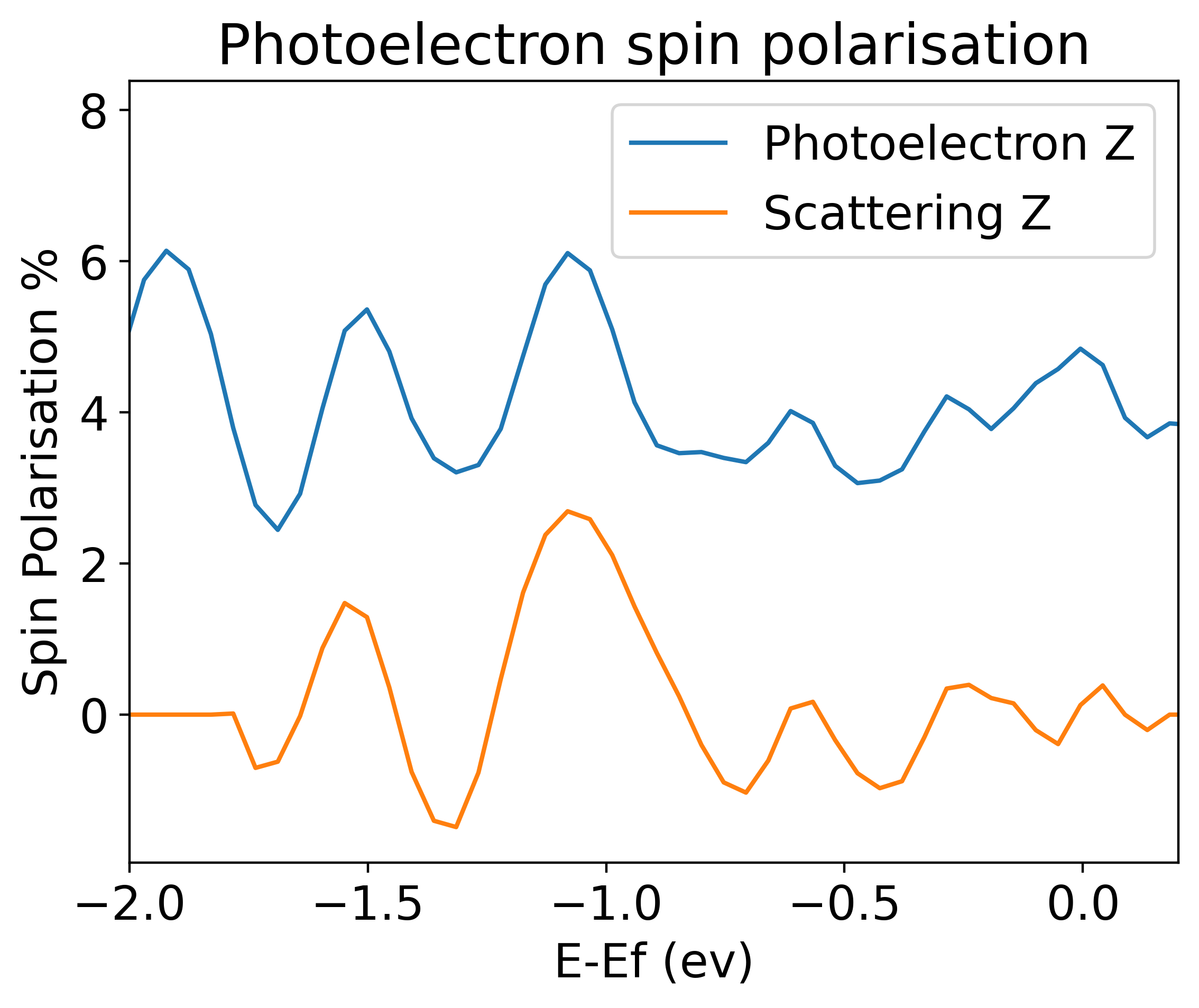}
    \caption{Energy resolved spin polarised current along $Z$ direction. Parameters are as shown in table \ref{table:Parameters_used} except for. Left: $\gamma_p^{\uparrow} = 1.04 \times 10^{-5}\Eh$, $\gamma_p^{\downarrow} = 0.96 \times 10^{-5}\Eh$ . Right: $\gamma_s^{\uparrow} = \gamma_s^{\downarrow}  = 10^{-3}\Eh$ }
    \label{SI:fig:SpinPolarisedCurrentSmallerGamma}
\end{figure}

The magnitude of the effect stays approximately constant as the parameters are varied. This is illustrated in Figure \ref{SI:fig:SpinPolarisedCurrentSmallerGamma} where $\gamma_p$ and $\gamma_s$ are reduced by factor $10$ and lead to qualitatively the same results. \\

\begin{figure}[ht!]
    \centering
    \includegraphics[width=0.45\linewidth]{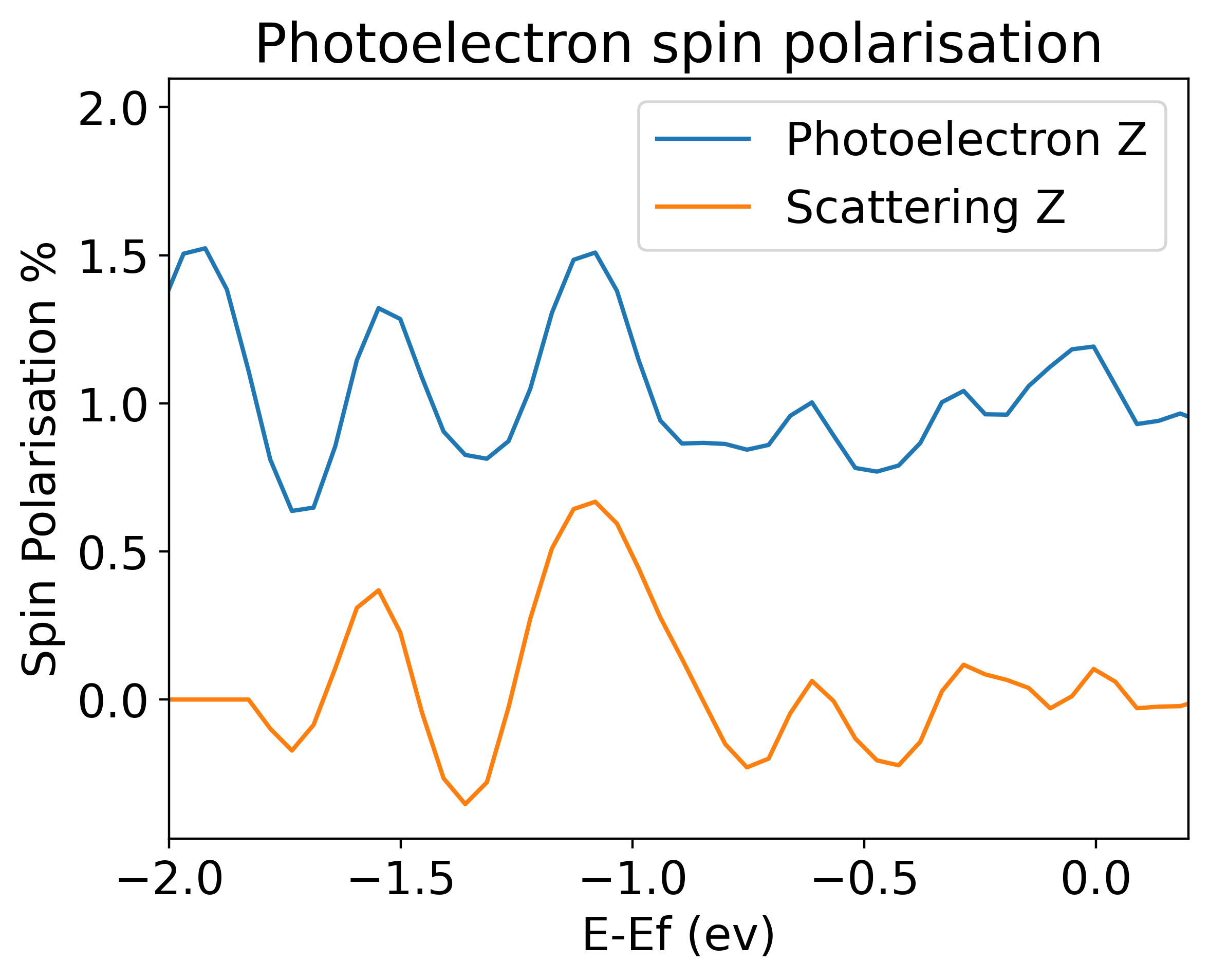}
    \includegraphics[width=0.45\linewidth]{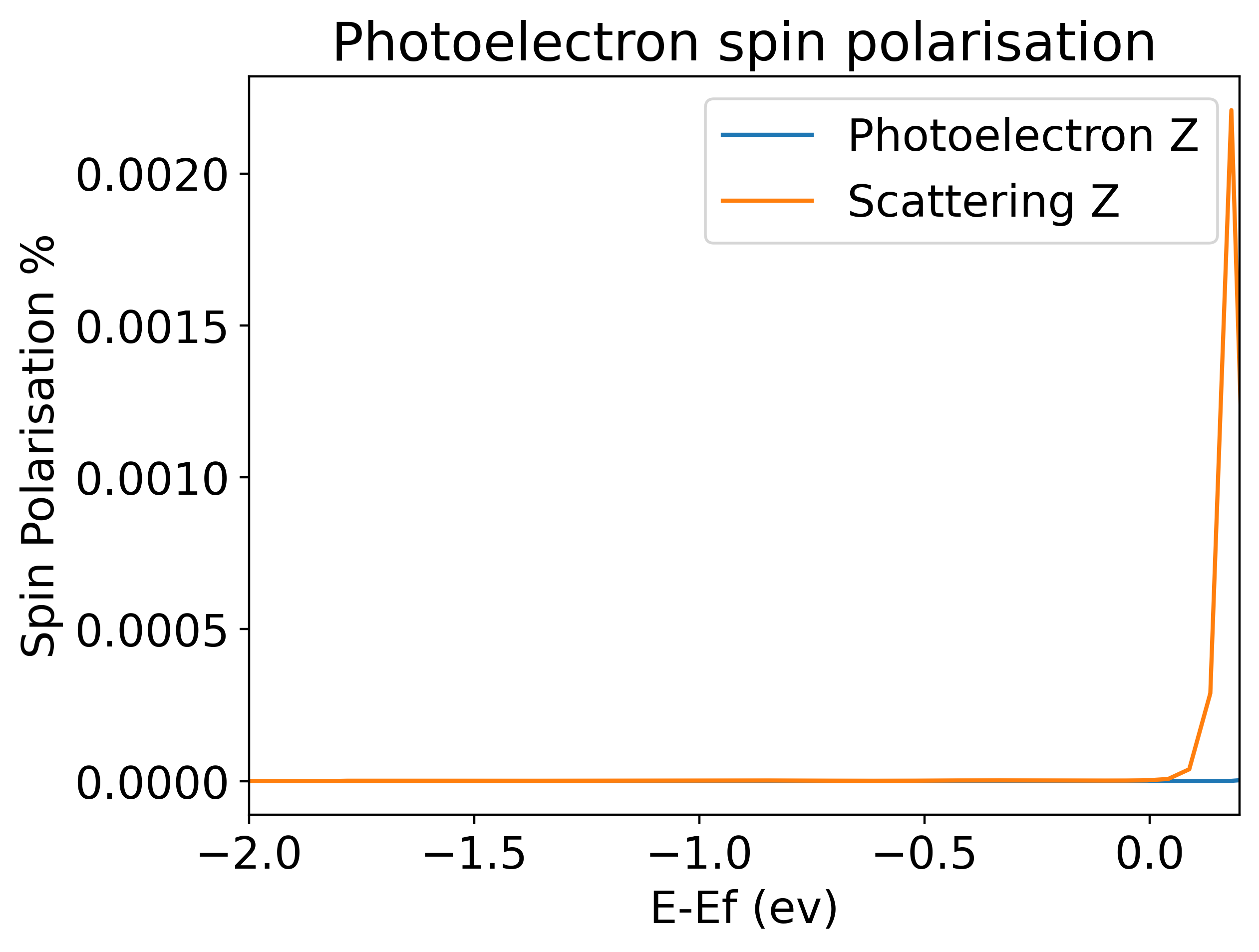}
    \caption{Energy resolved spin polarised current along $Z$ direction. Parameters are as shown in table \ref{table:Parameters_used} except for. Factor $1.04 \rightarrow 1.01$, Factor $0.96 \rightarrow 0.99$. Right: $1.04 \rightarrow 1$, Factor $0.96 \rightarrow 1$}
    \label{SI:fig:SpinPolarisedCurrentSmallerBias}
\end{figure}

The bulk copper spin bias, $\gamma_b$ can also be changed. Decreasing the initial spin bias, decreases the magnitude of the CISS effect. Figure \ref{SI:fig:SpinPolarisedCurrentSmallerBias} shows this for a spin bias of $1\%$ and $0\%$. At $0\%$ the spin symmetric solution is recovered. The slight apparent polarisation just above the Fermi level is due to numerical issues with dividing very small numbers.

\bibliographystyle{unsrt}
\bibliography{CISS}